\documentclass[b5paper]{report}
\usepackage{epsfig,amsmath,amscd,amssymb,amsfonts,calc,cite,dissertation,graphicx,indentfirst,subfigure,supertabular, color}
\usepackage{bigints}
\usepackage{makecell}
\usepackage{multirow}
\usepackage{algpseudocode,algorithm,algorithmicx}

\usepackage{tabularx}
\newcommand{\multiline}[1]{%
  \begin{tabularx}{\dimexpr\linewidth-\ALG@thistlm}[t]{@{}X@{}}
    #1
  \end{tabularx}
}
\algnewcommand{\IfThenElse}[3]{
  \State \algorithmicif\ #1\ \algorithmicthen\ #2\ \algorithmicelse\ #3}
\usepackage[toc,page]{appendix}
\fancy
\author{Hyunsoo Kim}
\title{Ultra-Reliable Urban Air Mobility Networks}
\newcommand{\referee}{Daesik Hong}
\newcommand{\secondreferee}{Chungyong Lee}
\newcommand{\thirdreferee}{Sooyong Choi}
\newcommand{\fourthreferee}{Hanho Wang}
\newcommand{\fifthreferee}{Wonsuk Chung}

\begin{document}
\pagenumbering{roman}
\maketitle
\pagestyle{empty}
\pagestyle{plain}
\setcounter{page}{1}
\tableofcontents

\listoffigures 
\listoftables
\chapter*{List of Abbreviations}
\addcontentsline{toc}{chapter}{List of Abbreviations}
\begin{supertabular}{ll}
UAM & Urban Air Mobility \\
C2 & Command and Control \\
VTOL & Vertical Take-Off and Landing \\
AGL & Above ground level \\
3GPP & 3rd Generation Partnership Project \\
KPI & Key Performance Indicator \\
UAV & Unmanned Aerial Vehicle \\
MBS & Macro Base Station \\
IMD & Inter-MBS Distance \\
SINR & Signal-to-Interference-plus-Noise Ratio \\
PPP & Poisson Point Process \\
SLS & System-Level Simulation \\
CDF & Cumulative Density Function \\
MGF & Moment Generating Function \\
PGFL & Probability Generating FunctionaL \\

\end{supertabular} \clearpage 
\pagenumbering{roman}
\setcounter{page}{8}
\begin{abstract}
\vspace{-3mm}
Recently, urban air mobility (UAM) has attracted attention as an emerging technology that will bring innovation to urban transportation and aviation systems. Since the UAM systems pursue fully autonomous flight without a pilot, wireless communication is a key function not only for flight control signals, but also for navigation and safety information. The UAM networks must be configured so that the UAM can receive command and control (C2) messages by securing continuous link stability without any interruptions. Nevertheless, a lot of prior works have focused only on improving the average performance without solving the low-reliability in the cell edges and coverage holes of urban areas.

In this dissertation, we identify the factors that hinder the communication link reliability in considering three-dimensional (3D) urban environments and propose an antenna configuration, resource utilization, and transmission strategy to enable UAM to receive C2 messages regardless of time and space. First, through stochastic geometry modeling, we analyze the signal blockage effects caused by the urban buildings. The blockage probability is calculated according to the shape, height, and density of the buildings, and the coverage probability of the received signal is derived by reflecting the blockage events. Furthermore, the low-reliability area is identified by analyzing the coverage performance according to the positions of the UAMs. To overcome the low-reliability region, we propose three methods for UAM network operation: i) optimization of antennas elevation tilting, ii) frequency reuse with multi-layered narrow beam, and iii) assistive transmission by the master UAM. To be specific, the optimal antenna tilting allow the C2 signal to be transmitted evenly to all regions, and the frequency reuse with multi-layered narrow beam is effective in reducing strong line of sight (LOS) interference at high altitudes. Lastly, the proposed master UAM transmission is designed to support UAMs located in the coverage hole.

Through system-level simulation (SLS), we evaluate the coverage and reliability performance of the proposed UAM network operation considering 3D UAM mobility. We demonstrate that the UAM network can improve the link reliability up to 99.9 \% when all three proposed operation methods are applied comprehensively.

\end{abstract}

\pagenumbering{arabic}
\chapter{Introduction}
\hspace{-5mm}\rule{135mm}{1pt}

{1.1 Background 

1.2 Motivation and Contribution\\}
\rule{135mm}{1pt}
\clearpage

\section{Background} \label{Sec_1.1}

Urban air mobility (UAM) is an emerging technology that will bring innovation to urban transportation and aviation systems. As an extension of research on autonomous systems, it is a technology that improves local accessibility to take passengers or cargoes into and out of the city by overcoming the constraints of time and space. In detail, use cases include last-mile delivery using drones, air metro for next-generation public transport, and air tax for door-to-door rides \cite{0}.

Since UAM systems pursue fully autonomous flight without a pilot, wireless communication is a key function not only for flight control signals, but also for navigation and safety information. The UAM system, which features vertical take-off and landing (VTOL), operates above ground level (AGL) that the conventional terrestrial communication system has not considered as a service area.

Table \ref{Table1_1} describes key performance indicators (KPIs) for autonomous UAM command and control (C2) in the 3rd Generation Partnership Project (3GPP) TS 22.125 \cite{1}. The KPI considers two categories: i) horizontal autonomous driving and ii) VTOL on infrastructure. It defines the maximum speed of a UAM for each situation, the transmission interval and size of the control message, and the required end-to-end latency. Reliability is defined as the probability of successful transmission within the required latency at the application layer while under network coverage and should achieve more than 99 \%. It is noted that the latency and message size requirements in a VTOL category directly related to safety are tighter than those in the autonomous flight category.

\clearpage

\begin{table}[t!]
    \centering
    \caption{KPIs for Command and Control of UAM Operation}\label{Table1_1}
    \includegraphics[width=1\columnwidth]{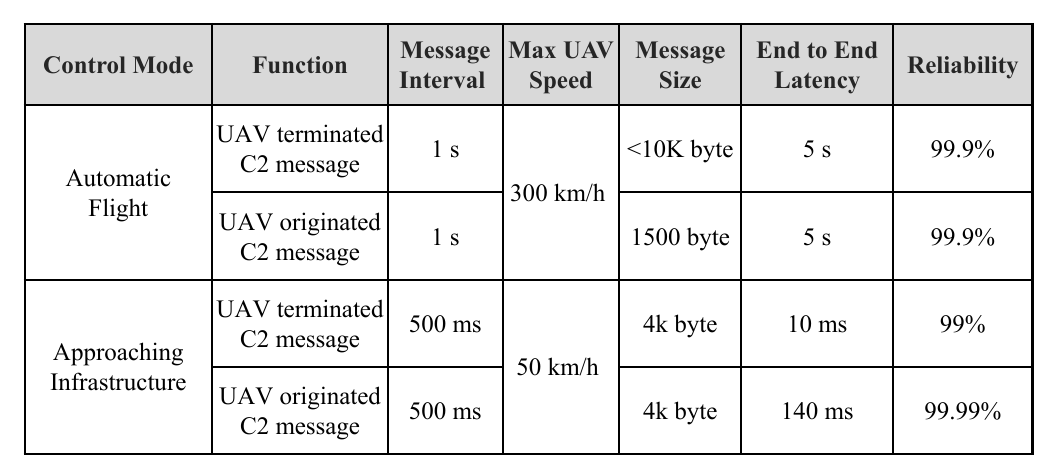}
\end{table}

Prior to discussing UAM systems, numerous studies have been actively conducted under the keyword unmanned aerial vehicle (UAV). While a UAM is recognized as transportation that requires continuous link reliability for passengers' safety, UAV is treated as a base station or a user equipment floating in the air. 

T. Izydorczyk et al. in \cite{3} designed a UAV test-bed and measured the quality of the received signal. In \cite{4, 5, 6}, the effect of the UAV on the ground cellular user was analyzed, and an interference cancellation technique at the ground user side was proposed. In \cite{7, 8 ,9}, ground users conduct non-orthogonal transmission to a UAV base station (UAV-BS). S. Zhang et al. formulated the sub-channel allocation and UAV speed optimization problem to maximize the uplink sum-rate in a single-cell cellular network where multiple UAVs upload their collected data to a single BS \cite{10}. In \cite{11, 12}, interference mitigation schemes were proposed via cooperative receiving at multiple BSs under strong line-of-sight (LOS) links. Cooperation among BSs is mandatory, which requires high complexity. In \cite{13, 14}, the throughput in two-hop relay system of ground-UAV-satellite was analyzed. However, it is assumed that ground, UAV, and satellite use the same frequency resource. This is an unfeasible system model that cannot achieve high reliability of UAVs by giving strong interference to each other's networks.

In \cite{15, 16, 17}, when the UAV collects data from ground nodes, a trajectory to minimize flight energy while maximizing the received data rate was studied. M. A. Ali et al. in \cite{18} proposed UAV placement and power allocation. S. Zhang et al. proposed a sequential cell association in non-uniform multi-cell environments \cite{19}. In \cite{20}, UAV-assisted internet of things (IoT) networks were investigated with IoT power optimization. A. Ranjha et al. proposed a three-dimensional (3D) cellular network where drone BSs serve drone users \cite{21}. Even though the UAV serves as a downlink base station, there is no consideration for a backhaul link for the UAV to provide reliable communication links. 

In \cite{22}, beamformers were designed to provide a certain time-invariant signal-to-interference-plus-noise ratio (SINR) to minimize the SINR fluctuation. However, in this paper, there is no consideration and solution for an air channel with strong LOS characteristics, which bring about strong interference to UAVs. W. Tang et al. proposed tractable modeling for UAV networks where multiple UAV-BSs serve ground users \cite{24}. In particular, the characteristics of the urban environment were discussed through the analysis of penetration loss in buildings. However, it is assumed that all UAVs have the same altitude, and there is no solution to improve link reliability, just for observation.

\clearpage

\section{Motivation and Contribution} \label{Sec_1.3}

A common limitation of prior works is that they focus only on improving average performance without considering the low-reliability in the cell edges and coverage holes caused by building blockages. However, in the UAM system, it is important to ensure the minimum performance required to reliably receive control signals from individual UAMs anytime and anywhere. The reliability of the UAM system means continuous link stability, not instantaneous signal-to-interference-plus-noise ratio (SINR) performance. Accordingly, it is necessary to study the network configuration and transmission technology that can guarantee link reliability under a practical 3D urban model.

In this dissertation, we consider a cellular-based downlink UAM network that supports ultra-reliable UAM operations. The final goal is to achieve 99.9 \% reliability without the loss of C2 messages. We investigate the geometry SINR performance of UAM systems with macro BS (MBS) and UAM deployment. Based on the performance analysis, we propose the MBS configuration and transmission techniques to improve link reliability. The key contributions in this dissertation are summarized as:

\begin{itemize}
    \item \textbf{Analysis of Link Characteristics in 3D Urban Area}
    
    \,\,\,\,We analyze the blockage effect in 3D urban model. Based on the Poisson point process (PPP), coverage probability is derived to discuss the characteristics of UAM networks. Especially, the channel characteristics of the VTOL zone, where signal blocking occurs frequently, and the horizontal flight zone, which is subjected to strong LOS interference, are observed.

\clearpage
    \item \textbf{Discussion of the MBS Deployment and Configuration}
    
    \,\,\,\,Given 3D urban modeling, we find out the optimal inter-MBS distance and elevation tilting angle of antennas to maximize the coverage probability at high altitudes. The proposed setup allows UAMs to receive high-quality C2 messages anywhere.
    
    \item \textbf{Improvement of Continuous Link Reliability}
    
    \,\,\,\,We propose a frequency reuse pattern with multi-layered narrow beams. By performing beam and resource allocation suitable for VTOL zone and horizontal flight zone, respectively, the signal quality of low-performance UAM is significantly improved. In addition, we introduce an assistive transmission technique of a master UAM to support UAMs in coverage holes.

\end{itemize}

An outline of this dissertation is given as follows:
Chapter 2 describes the system layout and blockage model in a 3D urban scenario. This chapter represents system notation and conditions for a blockage event.
Chapter 3 provides the performance analysis of UAM networks based on the PPP models.
Specifically, we analyze the average number of blockage buildings for communication links and derive the coverage probability of UAM networks with various system parameters. Furthermore, through system-level simulation (SLS), we verify the coverage probability performance and find out the cell edge issue at a high altitude to overcome for ultra-reliable UAM networks. 
In chapter 4, we investigate an optimal tilting angle for the antennas of UAM networks and the proposed frequency reuse pattern with multi-layered beams. In addition, we propose a master UAM algorithm for assistive transmissions to support UAMs which do not satisfy reliability requirements. Finally, chapter 5 presents the conclusions and topics for future research.

\chapter{System Model of Urban Air Mobility Communications}\label{Chap_2}

\hspace{-5mm}\rule{135mm}{1pt}

{2.1 System Layout 

2.2 Blockage Model in 3D urban Scenario 

\,\,\,\,\,\,\,\,\,2.2.1 Vertical Condition

\,\,\,\,\,\,\,\,\,2.2.2 Horizontal Condition\\}
\rule{135mm}{1pt}
\clearpage

\section{System Layout} \label{Sec_2.1}

\begin{figure}[p!]
    \centering
    \includegraphics[width=0.9\columnwidth]{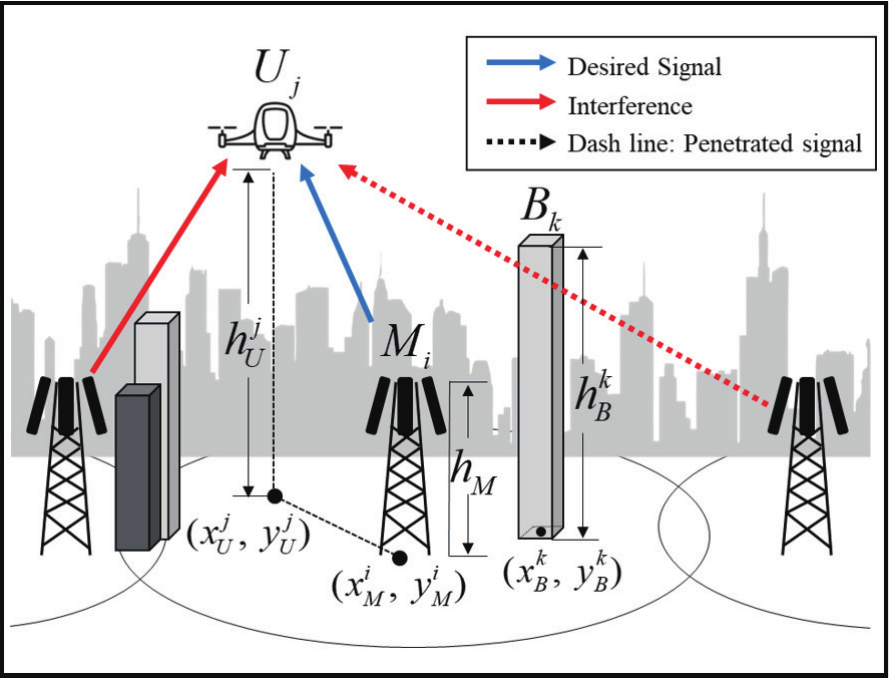}
    \caption{The system layout of UAM networks. $(x_M^i,\,\,y_M^i,\,h_M)$ and ${(x_U^j,\,\,y_U^j,\,h_U^j)}$, $(x_B^k,\,\,y_B^k,\,h_B^k)$ are the coordinates of the ${i_{th}}$ MBS ${M_i}$, ${j_{th}}$ UAM ${U_j}$, and the center point of the ${k_{th}}$ building top, respectively.}
    \label{Fig2_1}
\end{figure}

A 3D cellular-based downlink UAM networks in urban environment is considered as shown in Figure \ref{Fig2_1}. The distribution of the MBS follows a Poisson Point Process (PPP) ${\Phi _M}$ with density ${\lambda _M}$ MBSs/${{\rm{km}}^2}$, over which UAMs are positioned as independent PPP ${\Phi _U}$ with density ${\lambda _U}$ UAMs/${{\rm{km}}^2}$. MBSs and UAMs are denoted ${\Phi _M} = \left\{ {{M_1},\,{M_2},\, \cdots ,\,{M_{{N_M}}}} \right\}$ and ${\Phi _U} = \left\{ {{U_1},\,{U_2},\, \cdots ,\,{U_{{N_U}}}} \right\}$, where ${N_M}$ and ${N_U}$ are the number of MBSs and UAMs, respectively. We assume that the MBSs have a constant height ${h_M}$ and UAMs have a height distribution according to uniform random variable between ${h_U^{\min }}$ $m$ and ${h_U^{\max }}$ $m$, i.e., $h_U^j{\rm{ }} \sim {\rm{ uniform }}\left[ {h_U^{\min }{\rm{, }}\,h_U^{\max }} \right]$. Therefore, the coordinates of the ${i_{th}}$ MBS ${M_i}$ can be expressed as $(x_M^i,\,\,y_M^i,\,h_M)$ and the coordinates of the ${j_{th}}$ UAM ${U_j}$ can be expressed as ${(x_U^j,\,\,y_U^j,\,h_U^j)}$. The 2D distance in XY plane between ${i_{th}}$ MBS and ${j_{th}}$ UAM can be presented as $r_{UM}^{ji} = \sqrt {{{\left( {x_U^j - x_M^i} \right)}^2} + {{\left( {y_U^j - y_M^i} \right)}^2}}$.

The buildings are modeled as a square shape of width $W$ depth $D$, and the center point of the square follows the PPP ${\Phi _B} = \left\{ {{B_1},\,{B_2},\, \cdots ,\,{B_{{N_B}}}} \right\}$ where ${{N_B}}$ is the number of buildings. Buildings’ heights distribute according to Rayleigh distribution with probability density function
\begin{equation}\label{Bd_height_Distribution}
{f_{{h_B}}}(h_B^k) = \frac{{h_B^k}}{{{{\left( {{\sigma _{{h_B}}}} \right)}^2}}}\exp \left( { - \frac{{{{\left( {h_B^k} \right)}^2}}}{{2{{\left( {{\sigma _{{h_B}}}} \right)}^2}}}} \right)
\end{equation}
where ${{\sigma _{{h_B}}}}$ is the scale parameter of Rayleigh distribution, and the average height of buildings is $\sqrt {\frac{\pi }{2}} {\sigma _{{h_B}}}$ \cite{24}. The center point of the ${k_{th}}$ building top can be denoted by $(x_B^k,\,\,y_B^k,\,h_B^k)$
Denote $N_{B}^{ji}$ as the number of buildings that intersect link between ${i_{th}}$ MBS and ${j_{th}}$ UAM, which means the number of blockage events.

\clearpage

The total penetration loss is expressed by
\begin{equation}\label{PL}
    L_{UM}^{ji} = {\gamma ^{N_B^{ji}}},
\end{equation}
where $\gamma\left( {0 \le \gamma  \le 1} \right)$ is scale factor of penetration loss. The smaller $\gamma$ value is, the higher the penetration loss occurs. As a special case, $\gamma  = 0$ means that the signal is completely blocked \cite{25}.

In this system, it is assumed that the UAMs are associated with the MBS having the strongest reference signal received power (RSRP) value. Given the serving MBS ${M_{i'}}$, we can derive signal to interference ratio (SIR) of the $j_th$ UAM $U_j$ as
\begin{equation}\label{SINR}
{\rm{SINR(}}r_{UM}^{ji'},\,\,h_U^j{\rm{) = }}\frac{{{P_{TX}}G_{UM}^{ji'}L_{UM}^{ji'}{{\left( {{{\left( {r_{UM}^{ji'}} \right)}^2} + {{\left( {h_U^j - {h_M}} \right)}^2}} \right)}^{{\raise0.7ex\hbox{${ - \alpha }$} \!\mathord{\left/
 {\vphantom {{ - \alpha } 2}}\right.\kern-\nulldelimiterspace}
\!\lower0.7ex\hbox{$2$}}}}}}{{\sum\limits_{i \in {\Phi _M}{\rm{\backslash }}i'} {{P_{TX}}G_{UM}^{ji}L_{UM}^{ji}{{\left( {{{\left( {r_{UM}^{ji}} \right)}^2} + {{\left( {h_U^j - {h_M}} \right)}^2}} \right)}^{{\raise0.7ex\hbox{${ - \alpha }$} \!\mathord{\left/
 {\vphantom {{ - \alpha } 2}}\right.\kern-\nulldelimiterspace}
\!\lower0.7ex\hbox{$2$}}}} + {\sigma ^2}} }},
\end{equation}
where $x={r_{UM}^{ji'}}$ is the 2D horizontal distance from the $U_j$ to the ${M_{i'}}$. $P_{TX}$ is transmit power of MBSs and $\alpha$ is a path-loss exponent. The ground-to-air channel model is assume to be Nakagami-m fading channel of which power gain $G$ follows gamma distribution ${f_G}(g) = \frac{{{m^m}{g^{m - 1}}}}{{\Gamma (m)}}\exp ( - mg)$. ${{\sigma ^2}}$ is a noise power.

\clearpage

\section{Blockage Model in 3D Urban Scenario} \label{Sec_2.2}
In this section, we will consider the conditions of the blocking event required to derive a penetration loss.
There are two conditions: vertical one and horizontal one \cite{25}.

\subsection{Vertical Condition} \label{SubSec_2.2.1}
    
\begin{figure}[p!]
    \centering
    \includegraphics[width=0.9\columnwidth]{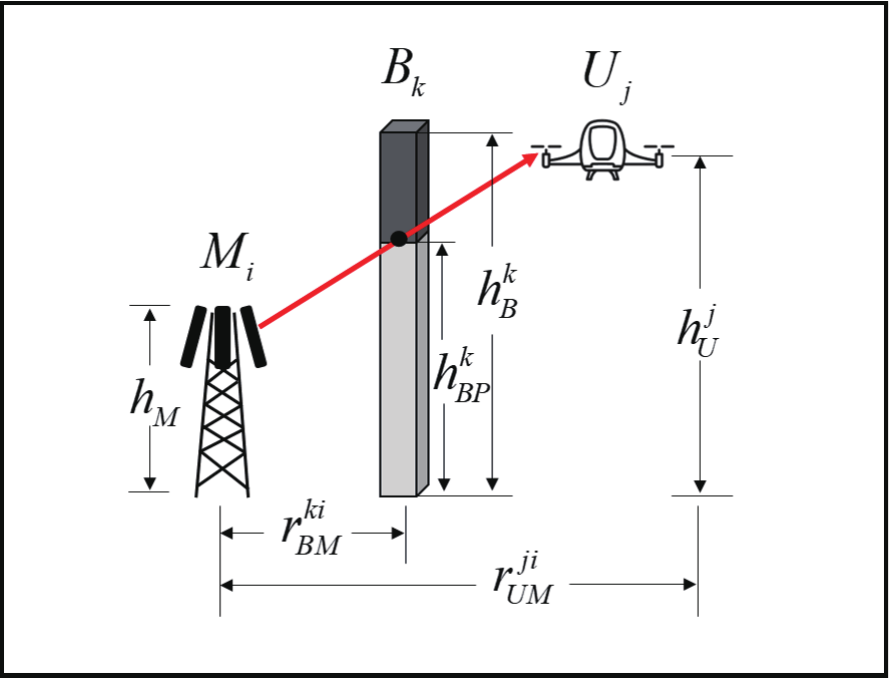}
    \caption{The vertical condition of blockage events. $h_{BP}^k$ is penetration point of the building. When $h_{BP}^k$ is lower than the building height $h_{B}^k$, the link $\overline {{M_i}{U_j}}$ experiences blockage event.}
    \label{Fig2_2}
\end{figure}

For simplicity of analysis, we focus on the height without the width and depth of the building.
    As shown in Figure \ref{Fig2_2}, assuming that the MBS, UAM, and building are all on the same straight line, the building intersects the direct propagation path from the MBS to the UAM, $\overline {{M_i}{U_j}}$ in vertical plane. In other words, the penetration point $h_{BP}^k$ of the building should be lower than the height $h_B^k$ of the building. By triangular proportional expression, the vertical condition can be represented as
\begin{equation} \label{Vertical_condition_1}
h_{BP}^k = \frac{{r_{BM}^{ki}\left( {h_U^j - {h_M}} \right)}}{{r_{UM}^{ji}}} + {h_M} \le h_B^k\,\,\,\,{\rm{for}}\,\,{h_M} < h_U^j.
 \end{equation}
 The penetration point of the building $h_{BP}^k$ is proportional to the distance between a MBS and a building, the height difference between a UAM and a MBS, and inversely proportional to the distance between a UAM and a MBS. The lower the penetration point, the more likely the blockage will occur.

In the same way, when the height of the UAM is lower than the height of the base station, the following conditions should be satisfied.
\begin{equation} \label{Vertical_condition_2}
h_{BP}^k = \frac{{(r_{UM}^{ji}-r_{BM}^{kj})\left( {{h_M} - h_U^j} \right)}}{{r_{UM}^{ji}}} + h_U^j \le h_B^k\,\,\,\,{\rm{for}}\,\,{h_M} > h_U^j.
\end{equation} 
Contrary to \eqref{Vertical_condition_1}, the further the distance between a building and a MBS, the more likely the blockage event occurs.

\subsection{Horizontal Condition} \label{SubSec_2.2.2}

Figure \ref{Fig2_2} describes the example of blockage condition in the first quadrant horizontal plane. Given the angle of the $\overline {{M_i}{U_j}}$ in horizontal plane, i.e., $\theta _{UM}^{ji}$, it needs to exist between the maximum and minimum angles of the straight line between the MBS and the vertices of each building, $\theta _{min}^{ji}$ and $\theta _{max}^{ji}$. The horizontal condition of the blockages can be expressed by
    
\begin{equation}\label{Horizontal_condition}
\begin{array}{c}
\theta _{min}^{ji} \le \theta _{UM}^{ji} = ta{n^{ - 1}}\left( {\frac{{y_U^j - y_M^i}}{{x_U^j - x_M^i}}} \right) \le \theta _{max}^{ji},\\
{\rm{where }}\begin{array}{*{20}{c}}
{\theta _{min}^{ji} = ta{n^{ - 1}}\left( {\frac{{\left( {y_U^j - \frac{D}{2}} \right) - y_M^i}}{{\left( {x_B^k + \frac{W}{2}} \right) - x_M^i}}} \right)}\\
{\theta _{max}^{ji} = ta{n^{ - 1}}\left( {\frac{{\left( {y_U^j + \frac{D}{2}} \right) - y_M^i}}{{\left( {x_B^k - \frac{W}{2}} \right) - x_M^i}}} \right)}
\end{array}for\,\,0 \le \theta _{UM}^{ji} \le \frac{\pi }{2}.
\end{array}
\end{equation}
The closer the center point of the building is to the MBS, the greater the difference between the maximum and minimum angles; as a result, the better the blockage event occurs. This shows a pattern similar to the vertical condition.

In the same way, the conditions in the $2_{nd}$, $3_{rd}$ and $4_{th}$ quadrants are as follows:
\begin{equation}\label{Horizontal_condition2}
\begin{array}{*{20}{c}}
{\theta _{min}^{ji} = ta{n^{ - 1}}\left( {\frac{{\left( {y_U^j + \frac{D}{2}} \right) - y_M^i}}{{\left( {x_B^k + \frac{W}{2}} \right) - x_M^i}}} \right)}\\
{\theta _{max}^{ji} = ta{n^{ - 1}}\left( {\frac{{\left( {y_U^j - \frac{D}{2}} \right) - y_M^i}}{{\left( {x_B^k - \frac{W}{2}} \right) - x_M^i}}} \right)}
\end{array}\,\,for\,\,\frac{\pi }{2} \le \theta _{UM}^{ji} \le \pi,
\end{equation}

\begin{equation}\label{Horizontal_condition3}
\begin{array}{*{20}{c}}
{\theta _{min}^{ji} = ta{n^{ - 1}}\left( {\frac{{\left( {y_U^j + \frac{D}{2}} \right) - y_M^i}}{{\left( {x_B^k - \frac{W}{2}} \right) - x_M^i}}} \right)}\\
{\theta _{max}^{ji} = ta{n^{ - 1}}\left( {\frac{{\left( {y_U^j - \frac{D}{2}} \right) - y_M^i}}{{\left( {x_B^k + \frac{W}{2}} \right) - x_M^i}}} \right)}
\end{array}\,\,for\,\, - \pi  \le \theta _{UM}^{ji} \le  - \frac{\pi }{2},
\end{equation}

\begin{equation}\label{Horizontal_condition4}
\begin{array}{*{20}{c}}
{\theta _{min}^{ji} = ta{n^{ - 1}}\left( {\frac{{\left( {y_U^j - \frac{D}{2}} \right) - y_M^i}}{{\left( {x_B^k - \frac{W}{2}} \right) - x_M^i}}} \right)}\\
{\theta _{max}^{ji} = ta{n^{ - 1}}\left( {\frac{{\left( {y_U^j + \frac{D}{2}} \right) - y_M^i}}{{\left( {x_B^k + \frac{W}{2}} \right) - x_M^i}}} \right)}
\end{array}\,\,for\,\, - \frac{\pi }{2} \le \theta _{UM}^{ji} \le 0.
\end{equation}

\begin{figure}[p!]
    \centering
    \includegraphics[width=0.9\columnwidth]{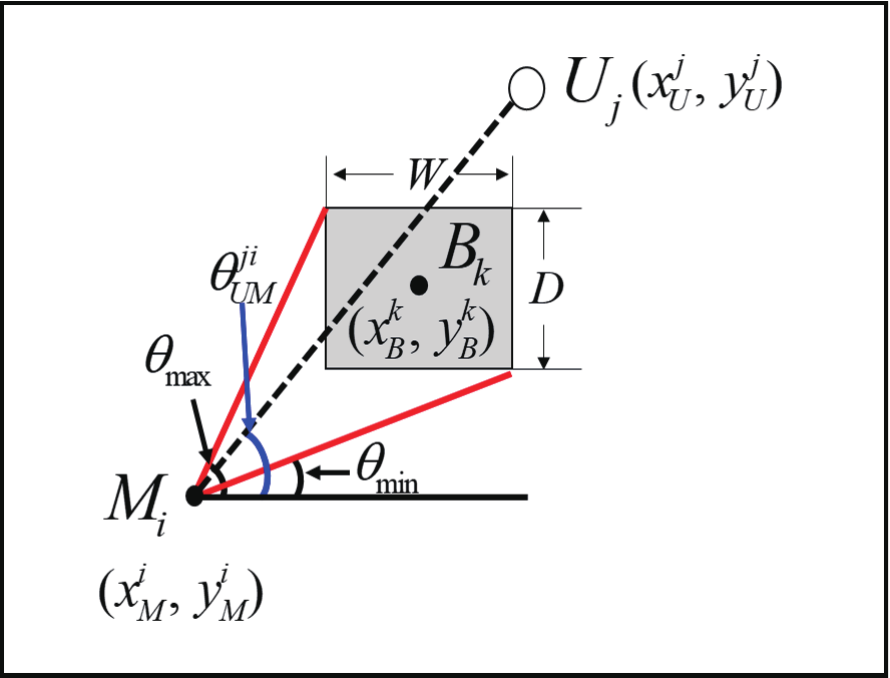}
    \caption{The horizontal condition of blockage event. $\theta _{UM}^{ji}$ is the angle of the $\overline {{M_i}{U_j}}$ in horizontal plane. When $\theta _{UM}^{ji}$ is positioned between the maximum and minimum angles of the straight line between the MBS and the vertices of each building, the link $\overline {{M_i}{U_j}}$ experiences blockage event.}
    \label{Fig2_3}
\end{figure}

\begin{table}[p!]
\begin{center}
\caption{Notation and description}\label{Table2_1}
\includegraphics[width=0.9\columnwidth]{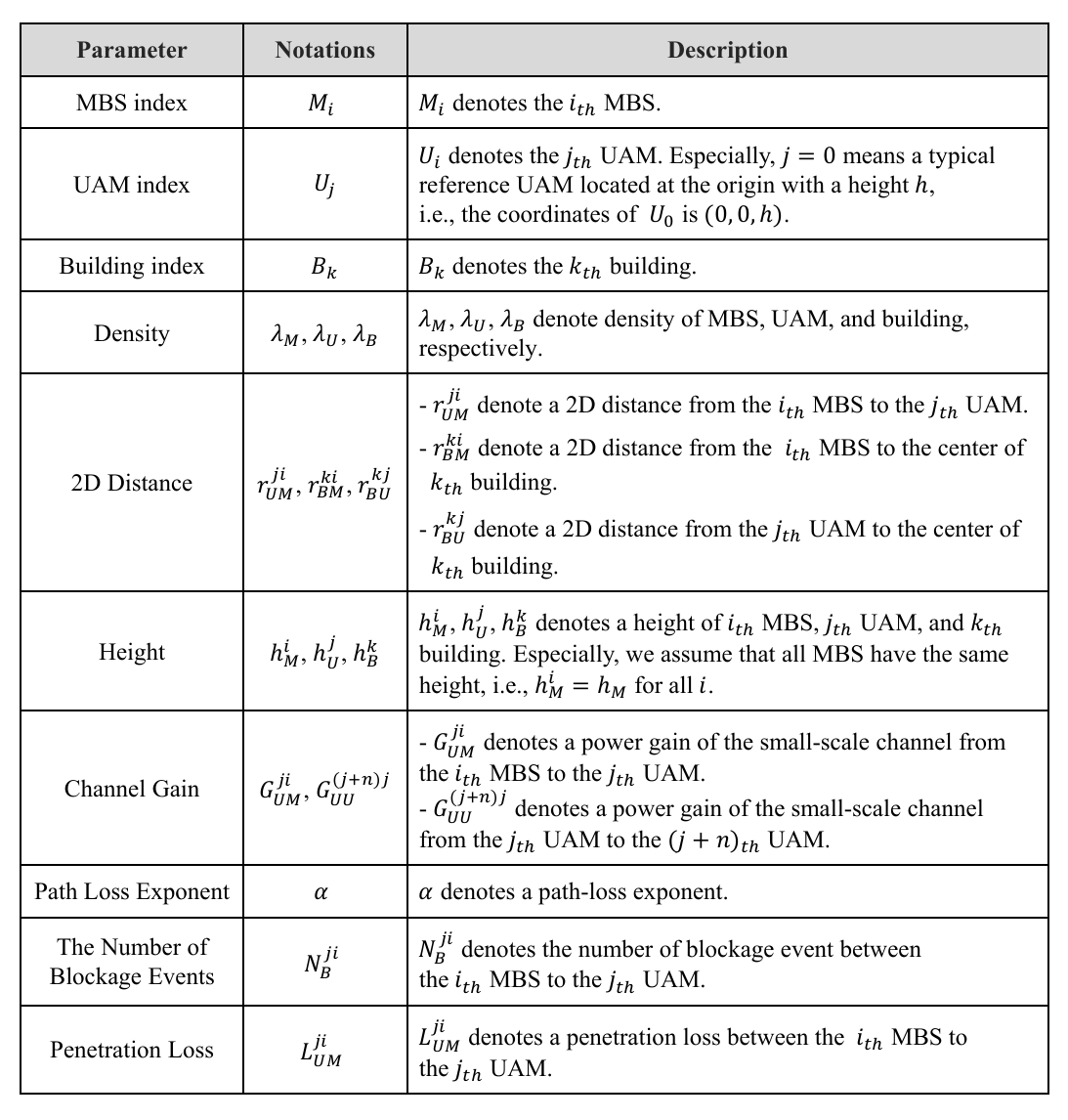}
\end{center}
\end{table}
\chapter{Penetration Loss and Coverage Probability Analysis}\label{Chap_3}
\hspace{-5mm}\rule{135mm}{1pt}

{3.1 Penetration Loss Analysis 

\,\,\,\,\,\,\,\,\,3.1.1 Probability of Blockage Event in Vertical Plane

\,\,\,\,\,\,\,\,\,3.1.2 The Average Number of Blockage Buildings in Horizontal Plane

\,\,\,\,\,\,\,\,\,3.1.3 Total Penetration Loss

3.2 Coverage Probability Analysis 

3.3 Numerical Results\\}
\rule{135mm}{1pt}
\clearpage

In this chapter, we derive the penetration loss and coverage probability based on the PPP model described in the Chapter 2. Especially, by analyzing the link quality according to the density of the building, the distribution of MBS, and the height of the UAM, we identify in which area the communication is weak. Note that, for a tractable analysis, we focus on a reference UAM $U_0$ with coordinates $(0, 0, h_U)$ in this chapter.

\section{Penetration Loss Analysis} \label{Sec_3.1}

\subsection{Probability of Blockage Event in Vertical Plane} \label{SubSec_3.1.1}
    
Given $r_{UM}$ and $h_{U}$, the probability that the penetration point of the building is lower than the height of the building is as follows:

\begin{equation}\label{total_Vertical_P}
\begin{array}{l}
{P_V}({r_{UM}},\,{h_U}) = \frac{1}{{{r_{UM}}}}\int\limits_{{r_{BM}} = 0}^{{r_{UM}}} {P\left( {{h_B} \ge {h_{BP}}} \right)d{r_{BM}}} \\
\mathop  = \limits^{(a)} \left\{ {\begin{array}{*{20}{c}}
{\frac{1}{{{r_{UM}}}}\int\limits_{r = 0}^{{r_{UM}}} {P\left( {{h_B} \ge h\frac{{({r_{UM}} - {r_{BM}})\left( {{h_M} - {h_U}} \right)}}{{{r_{UM}}}} + {h_U}} \right)d{r_{BM}}\,\,for\,\,\,{h_U} \le {h_M}} }\\
{\frac{1}{{{r_{UM}}}}\int\limits_{r = 0}^{{r_{UM}}} {P\left( {{h_B} \ge \frac{{{r_{BM}}\left( {{h_U} - {h_M}} \right)}}{{{r_{UM}}}} + {h_M}} \right)d{r_{BM}}} \,\,\,\,\,\,\,\,\,\,\,\,\,\,\,\,\,\,\,\,for\,\,\,{h_U} > {h_M}}
\end{array}} \right.\\
\mathop  = \limits^{(b)} \left\{ {\begin{array}{*{20}{c}}
{\frac{1}{{{r_{UM}}}}\int\limits_{r = 0}^{{r_{UM}}} {\exp \left( { - \frac{{{{\left( {\frac{{({r_{UM}} - {r_{BM}})\left( {{h_M} - {h_U}} \right)}}{{{r_{UM}}}} + {h_U}} \right)}^2}}}{{2{{\left( {{\sigma _{{h_B}}}} \right)}^2}}}} \right)d{r_{BM}}\,\,for\,\,\,{h_U} \le {h_M}} }\\
{\frac{1}{{{r_{UM}}}}\int\limits_{r = 0}^{{r_{UM}}} {\exp \left( { - \frac{{{{\left( {\frac{{{r_{BM}}\left( {{h_U} - {h_M}} \right)}}{{{r_{UM}}}} + {h_M}} \right)}^2}}}{{2{{\left( {{\sigma _{{h_B}}}} \right)}^2}}}} \right)d{r_{BM}}} \,\,\,\,\,\,\,\,\,\,\,\,\,\,\,\,\,for\,\,\,{h_U} > {h_M}}.
\end{array}} \right.
\end{array}
\end{equation}
where (a) follows vertical conditions based on \eqref{Vertical_condition_1}, \eqref{Vertical_condition_2}, and (b) follows the CDF of building heights, \eqref{Bd_height_Distribution}.

\clearpage

\subsection{The Average Number of Blockage Buildings in Horizontal Plane}\label{SubSec_3.1.2}

\begin{figure}[p!]
    \centering
    \includegraphics[width=0.8\columnwidth]{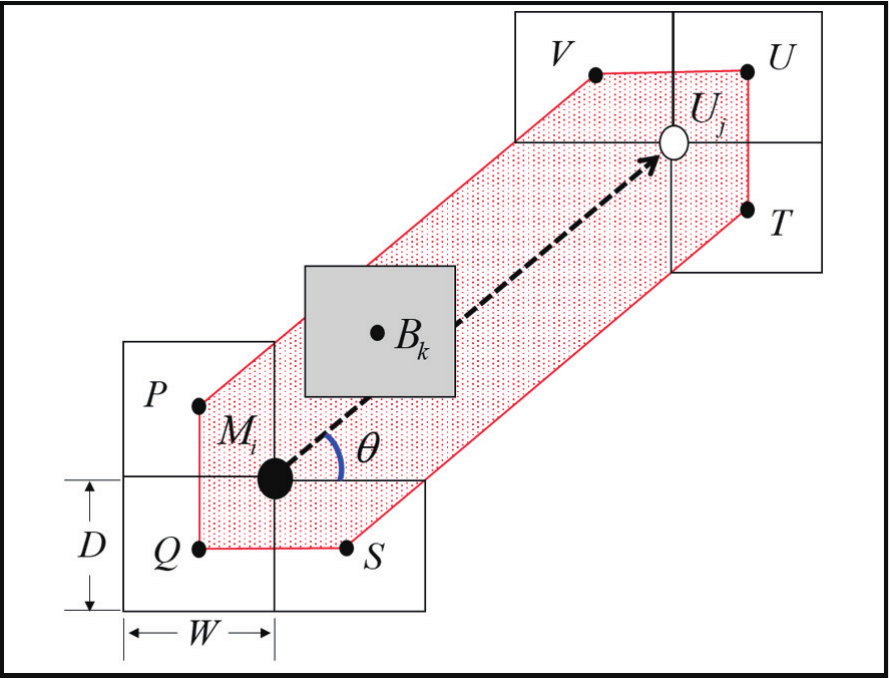}
    \caption{The region of which dropping points of the center of buildings in blockage events. When the the center of buildings is located in $PQSTUV$, the link $\overline {{M_i}{U_j}}$ experiences blockage event [25]. }
    \label{Fig3_1}
\end{figure}

Figure \ref{Fig3_1} shows the region of which the dropping points of the building center to occur blockage events. For a given $r_{UM}$, if the center point of the building is located within the PQSTUV, a blockage event occurs. Given $r_{UM}$ and $\theta_{UM}$, the area of PQSTUV can be obtained as 
\begin{equation}\label{Area}
A\left( {{r_{UM}}} \right) = {r_{UM}}W\left| {sin\theta } \right| + {r_{UM}}D\left| {cos\theta } \right| + WD,
\end{equation}
where $W$ and $D$ are the width and depth of the buildings, respectively. The longer the UAM-MBS distance and the larger the building size are, the wider the blockage area is \cite{25}.

By multiplying the blockage area by the density of the building, given $M_i$ and $U_j$, we can derive the number of blockage buildings as 
\begin{equation}\label{N}
{N_A}\left( {{r_{UM}}} \right) = {\lambda _B}\left( {{r_{UM}}W\left| {sin\theta } \right| + {r_{UM}}D\left| {cos\theta } \right| + WD} \right).
\end{equation}

Finally, given ${r_{UM}}$, the average number of blockage buildings in horizontal plane can be calculated as
\begin{equation}\label{Average_N}
\begin{array}{l}
\begin{array}{l}
E\left[ {{N_N}\left( {{r_{UM}}} \right)} \right] = \int\limits_\theta  {{\lambda _B}\left( {{r_{UM}}W\left| {sin\theta } \right| + {r_{UM}}D\left| {cos\theta } \right| + WD} \right)\frac{1}{{2\pi }}d\theta } \\
\,\,\,\,\,\,\,\,\,\,\,\,\,\,\,\, = \frac{{2{\lambda _B}\left( {W + D} \right)}}{\pi }{r_{UM}} + {\lambda _B}WD.
\end{array}
\end{array}
\end{equation}
The average number of blockage buildings in the horizontal plane is linearly proportional to UAM-MBS distance, building density, and the size of a building.

\clearpage

\subsection{Total Penetration Loss}\label{SubSec_3.1.3}

Since the probability of blockage in vertical plane and the average number of buildings in the horizontal plane are independent, from \eqref{PL}, \eqref{total_Vertical_P}, and \eqref{Average_N}, the total penetration loss can be obtained as follows:
\begin{equation}\label{PL_final}
\begin{array}{l}
\begin{array}{*{20}{c}}
{{L_{UM}} = {\gamma ^{E\left[ {{N_{BP}}\left( {{r_{UM}},\,{h_U}} \right)} \right]}},}\\
{{\rm{where}}\,\,E\left[ {{N_{BP}}\left( {{r_{UM}},\,{h_U}} \right)} \right] = {P_V}\left( {{r_{UM}},\,{h_U}} \right) \cdot E\left[ {{N_A}\left( {{r_{UM}}} \right)} \right].}
\end{array}\\
\end{array}
\end{equation}

\,\,\,\,As a special case $\gamma=0$, penetration loss can be Bernoulli random variable with probability
\begin{equation}\label{PL_Bernoulli}
\begin{array}{l}
P\left( {{L_{UM}} = 0} \right) = 1 - {e^{ - E\left[ {{N_{BP}}\left( {{r_{UM}},\,{h_U}} \right)} \right]}}\\
P\left( {{L_{UM}} = 1} \right) = {e^{ - E\left[ {{N_{BP}}\left( {{r_{UM}},\,{h_U}} \right)} \right]}},
\end{array}
\end{equation}
which means that signal is impenetrable to buildings \cite{24}.

\clearpage
\section{Coverage Probability Analysis} \label{Sec_3.2}
For simplicity of analysis, we assume UAM networks as interference-limited environment. From \eqref{SINR}, the coverage probability of reference UAM $U_0$ can be expressed by
\begin{equation} \label{Coverage_main}
\begin{array}{l}
{P_c}(T) = P\left( {SIR > T} \right)\\
\,\,\,\,\,\,\,\,\,\,\,\,\,\, = {E_{{r_{UM}},\,{h_U}}}\left[ {P\left( {{\rm{SIR(}}{r_{UM}},\,\,{h_U}{\rm{)}} > T} \right)} \right]\\
\,\,\,\,\,\,\,\,\,\,\,\,\,\, = {E_{{r_{UM}},\,{h_U}}}\left[ {P\left( {\frac{{{P_{TX}}{G_{UM}}{L_{UM}}{{\left( {{{\left( {{r_{UM}}} \right)}^2} + {{\left( {{h_U} - {h_M}} \right)}^2}} \right)}^{{\raise0.7ex\hbox{${ - \alpha }$} \!\mathord{\left/
 {\vphantom {{ - \alpha } 2}}\right.\kern-\nulldelimiterspace}
\!\lower0.7ex\hbox{$2$}}}}}}{I} > T} \right)} \right]\\
\,\,\,\,\,\,\,\,\,\,\,\,\,\, = \int\limits_{{h_U} = {h_U^{\min }}}^{{h_U^{\max }}} {\int\limits_{{r_{UM}} = 0}^\infty  {P\left[ {{G_{UM}} > T{P_{TX}}{L_{UM}}{{\left( {{{\left( {{r_{UM}}} \right)}^2} + {{\left( {{h_U} - {h_M}} \right)}^2}} \right)}^{{\raise0.7ex\hbox{$\alpha $} \!\mathord{\left/
 {\vphantom {\alpha  2}}\right.\kern-\nulldelimiterspace}
\!\lower0.7ex\hbox{$2$}}}}I} \right]} } \\
\,\,\,\,\,\,\,\,\,\,\,\,\,\,\,\,\,\,\,\,\,\,\,\,\,\,\,\,\,\,\,\,\,\,\,\,\,\,\,\,\,\,\,\,\,\,\,\, \times \,{f_{{r_{UM}}}}\left( {{r_{UM}}} \right){f_{{h_U}}}\left( {{h_U}} \right)d{r_{UM}}d{h_U},
\end{array}
\end{equation}
where $I = \sum\limits_{i \in {\Phi _M}{\rm{\backslash }}0} {{P_{TX}}G_{UM}^{ji}L_{UM}^{ji}{{\left( {{{\left( {r_{UM}^{ji}} \right)}^2} + {{\left( {h_U^j - {h_M}} \right)}^2}} \right)}^{{\raise0.7ex\hbox{${ - \alpha }$} \!\mathord{\left/
 {\vphantom {{ - \alpha } 2}}\right.\kern-\nulldelimiterspace}
\!\lower0.7ex\hbox{$2$}}}}}$ is the cumulative interference from all the other MBSs. ${f_{{r_{UM}}}}\left( {{r_{UM}}} \right) = {e^{{\lambda _M}\pi {r^2}}}2\pi {\lambda _M}r$ is the PDF of the 2D distance from a reference UAM to the closest MBS \cite{26}. ${f_{{h_U}}}\left( {{h_U}} \right) = {1 \mathord{\left/
 {\vphantom {1 {\left( {h_U^{\max } - h_U^{\min }} \right)}}} \right.
 \kern-\nulldelimiterspace} {\left( {h_U^{\max } - h_U^{\min }} \right)}}$ is the uniform distribution of UAM height.

As we calculate the middle part of \eqref{Coverage_main},
\begin{equation} \label{Coverage_1}
\begin{array}{l}
P\left[ {{G_{UM}} > T{P_{TX}}{L_{UM}}{{\left( {{{\left( {{r_{UM}}} \right)}^2} + {{\left( {{h_U} - {h_M}} \right)}^2}} \right)}^{{\raise0.7ex\hbox{$\alpha $} \!\mathord{\left/
 {\vphantom {\alpha  2}}\right.\kern-\nulldelimiterspace}
\!\lower0.7ex\hbox{$2$}}}}I} \right]\\
\mathop  = \limits^{(a)} {E_I}\left[ {{G_{UM}} > {{\Gamma \left( {m,\,mT{P_{TX}}{L_{UM}}{{\left( {{{\left( {{r_{UM}}} \right)}^2} + {{\left( {{h_U} - {h_M}} \right)}^2}} \right)}^{{\raise0.7ex\hbox{$\alpha $} \!\mathord{\left/
 {\vphantom {\alpha  2}}\right.\kern-\nulldelimiterspace}
\!\lower0.7ex\hbox{$2$}}}}I} \right)} \mathord{\left/
 {\vphantom {{\Gamma \left( {m,\,mT{P_{TX}}{L_{UM}}{{\left( {{{\left( {{r_{UM}}} \right)}^2} + {{\left( {{h_U} - {h_M}} \right)}^2}} \right)}^{{\raise0.7ex\hbox{$\alpha $} \!\mathord{\left/
 {\vphantom {\alpha  2}}\right.\kern-\nulldelimiterspace}
\!\lower0.7ex\hbox{$2$}}}}I} \right)} {\Gamma \left( m \right)}}} \right.
 \kern-\nulldelimiterspace} {\Gamma \left( m \right)}}} \right]\\
\mathop  = \limits^{(b)} {E_I}{\left[ {\sum\limits_{p = 0}^{m - 1} {\frac{{{{\left( {sI} \right)}^p}}}{{p!}}{e^{ - sI}}} } \right]_{s = mT{P_{TX}}{L_{UM}}{{\left( {{{\left( {{r_{UM}}} \right)}^2} + {{\left( {{h_U} - {h_M}} \right)}^2}} \right)}^{{\raise0.7ex\hbox{$\alpha $} \!\mathord{\left/
 {\vphantom {\alpha  2}}\right.\kern-\nulldelimiterspace}
\!\lower0.7ex\hbox{$2$}}}}}}\\
\mathop  = \limits^{(c)} \sum\limits_{p = 0}^{m - 1} {\frac{{{{\left( { - s} \right)}^p}}}{{p!}}\left[ {\frac{{{\partial ^p}}}{{\partial {s^p}}}{L_I}\left( {s\left| {{r_{UM}},\,} \right.\,{h_U}} \right)} \right]} ,
\end{array}
\end{equation}
(a) follows the CCDF of gamma distribution ${F_G}\left( {m,\,\,x} \right) = {{\Gamma \left( {m,\,x} \right)} \mathord{\left/
 {\vphantom {{\Gamma \left( {m,\,x} \right)} {\Gamma \left( m \right)}}} \right.
 \kern-\nulldelimiterspace} {\Gamma \left( m \right)}}$, (b) follows the regularized Gamma functions for Poisson random variables ${{\Gamma \left( {m,\,x} \right)} \mathord{\left/
 {\vphantom {{\Gamma \left( {m,\,x} \right)} {\Gamma \left( m \right)}}} \right.
 \kern-\nulldelimiterspace} {\Gamma \left( m \right)}} = \sum\limits_{p = 0}^{m - 1} {\frac{{{x^p}}}{{p!}}{e^{ - x}}} $, (c) can be represented as $p_{th}$ derivative of Laplace transform of $I$.
 
Assuming $\gamma = 1$ as a special case, the Laplace transform of $I$ is derived as
\begin{equation}\label{Laplace}
\begin{array}{l}
{L_I}\left( {s\left| {{r_{UM}},\,{h_U}} \right.} \right) = E\left[ {\exp \left( { - s\sum\limits_{r_{UM}^{ji} > x} {{P_{TX}}G_{UM}^{ji}L_{UM}^{ji}{{\left( {r_{UM}^{ji}} \right)}^{{\raise0.7ex\hbox{${ - \alpha }$} \!\mathord{\left/
 {\vphantom {{ - \alpha } 2}}\right.\kern-\nulldelimiterspace}
\!\lower0.7ex\hbox{$2$}}}}} } \right)} \right]\\
 = {\rm{ }}E\left[ {\prod\limits_{r_{UM}^{ji} > {r_{UM}}} {{E_{G_{UM}^{ji},\,L_{UM}^{ji}}}} \left[ {\exp \left( { - s{P_{TX}}G_{UM}^{ji}L_{UM}^{ji}{{\left( {r_{UM}^{ji}} \right)}^{{\raise0.7ex\hbox{${ - \alpha }$} \!\mathord{\left/
 {\vphantom {{ - \alpha } 2}}\right.\kern-\nulldelimiterspace}
\!\lower0.7ex\hbox{$2$}}}}} \right)} \right]} \right]\\
\mathop  = \limits^{(a)} {\rm{ }}E\left[ {\prod\limits_{r_{UM}^{ji} > {r_{UM}}} {{E_{G_{UM}^{ji}}}} } \right.\left[ {1 - {e^{{ - ^{E\left[ {{N_{BP}}\left( {{r_{UM}},\,{h_U}} \right)} \right]}}}}} \right.\\
\,\,\,\,\,\,\,\,\,\left. {\left. {\,\,\,\, + \exp \left( { - s{P_{TX}}G_{UM}^{ji}{{\left( {r_{UM}^{ji}} \right)}^{{\raise0.7ex\hbox{${ - \alpha }$} \!\mathord{\left/
 {\vphantom {{ - \alpha } 2}}\right.\kern-\nulldelimiterspace}
\!\lower0.7ex\hbox{$2$}}}}} \right){e^{{ - ^{E\left[ {{N_{BP}}\left( {{r_{UM}},\,{h_U}} \right)} \right]}}}}} \right]\,} \right]\\
\mathop  = \limits^{(b)} {\rm{ }}E\left[ {\prod\limits_{r_{UM}^{ji} > x} {1 - {e^{{ - ^{E\left[ {{N_{BP}}\left( {{r_{UM}},\,{h_U}} \right)} \right]}}}}} \left[ {1 - {{\left( {1 + \frac{{s{{\left( {r_{UM}^{ji}} \right)}^{{\raise0.7ex\hbox{${ - \alpha }$} \!\mathord{\left/
 {\vphantom {{ - \alpha } 2}}\right.\kern-\nulldelimiterspace}
\!\lower0.7ex\hbox{$2$}}}}}}{m}} \right)}^{ - m}}} \right]} \right]\\
\mathop  = \limits^{(c)} {\rm{ }}\exp \left[ { - 2\pi {\lambda _M}\int\limits_{{r_{UM}}}^\infty  {{e^{{ - ^{E\left[ {{N_{BP}}\left( {{r_{UM}},\,{h_U}} \right)} \right]}}}}} \left[ {1 - {{\left( {1 + \frac{{s{{\left( {r_{UM}^{ji}} \right)}^{{\raise0.7ex\hbox{${ - \alpha }$} \!\mathord{\left/
 {\vphantom {{ - \alpha } 2}}\right.\kern-\nulldelimiterspace}
\!\lower0.7ex\hbox{$2$}}}}}}{m}} \right)}^{ - m}}} \right]tdt} \right]
\end{array}
\end{equation}
where (a) follows Bernoulli random variable from \eqref{PL_Bernoulli}, (b) follows moment generating function (MGF) of Gamma distribution, (c) follows the probability generating functional (PGFL).

As inserting \eqref{Coverage_1} and \eqref{Laplace} into \eqref{Coverage_main}, the final result of coverage probability can be expressed by
\begin{equation}\label{Coverage_result}
\begin{array}{l}
{P_c}(T) = \int\limits_{{h_U} = 1.5}^{300} {\int\limits_{{r_{UM}} = 0}^\infty  {\sum\limits_{p = 0}^{m - 1} {\frac{{{{\left( { - s} \right)}^p}}}{{p!}}\left[ {\frac{{{\partial ^p}}}{{\partial {s^p}}}{L_I}\left( {s\left| {{r_{UM}},\,} \right.\,{h_U}} \right)} \right]} } } \\
\,\,\,\,\,\,\,\,\,\,\,\,\,\,\,\,\,\,\,\,\,\,\,\,\,\,\,\,\,\,\,\,\,\,\,\,\,\, \times {f_{{r_{UM}}}}({r_{UM}}){f_{{h_U}}}({h_U})d{r_{UM}}d{h_U}
\end{array}
\end{equation}

\clearpage

\section{Numerical Results} \label{Sec_3.3}

\begin{figure}[p!]
    \centering
    \includegraphics[width=0.7\columnwidth]{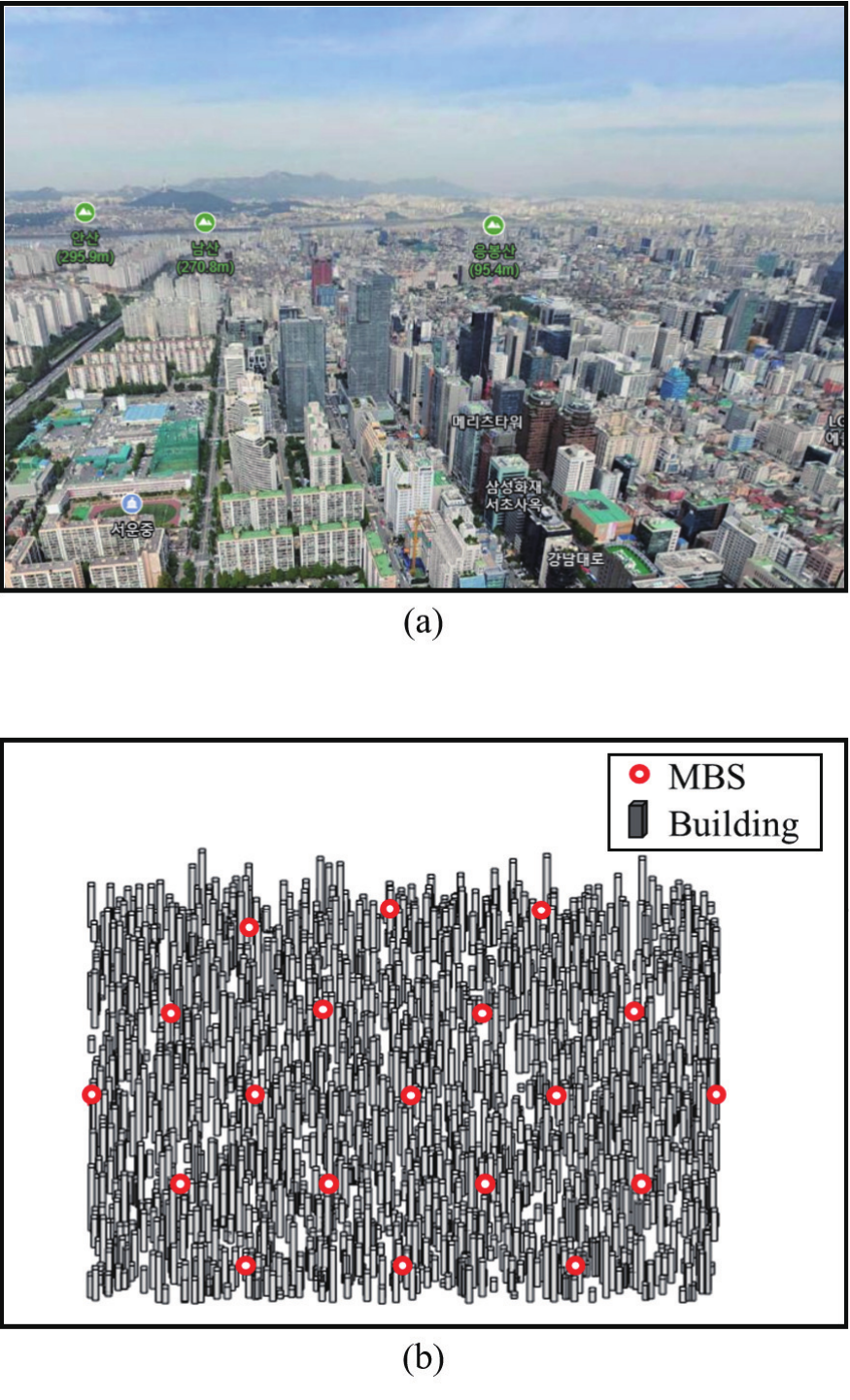}
    \caption{The simulation layout: (a) an aerial photography of Gang-Nam Station, (b) the designed system-level simulator}
    \label{Fig3_2}
\end{figure}

In this section, the performance of the UAM network is evaluated at the system-level. As depicted in Figure \ref{Fig3_2}, the simulator is designed taking into account the average building density and height in the area of the Gang-Nam Staten, South Korea. Based on 3GPP standard \cite{27, 28}, the specific simulation parameters are summarized in Table \ref{Table3_1}.

\begin{table}[h!]
\begin{center}
\caption{Simulation parameters.}\label{Table3_1}
\includegraphics[width=0.9\columnwidth]{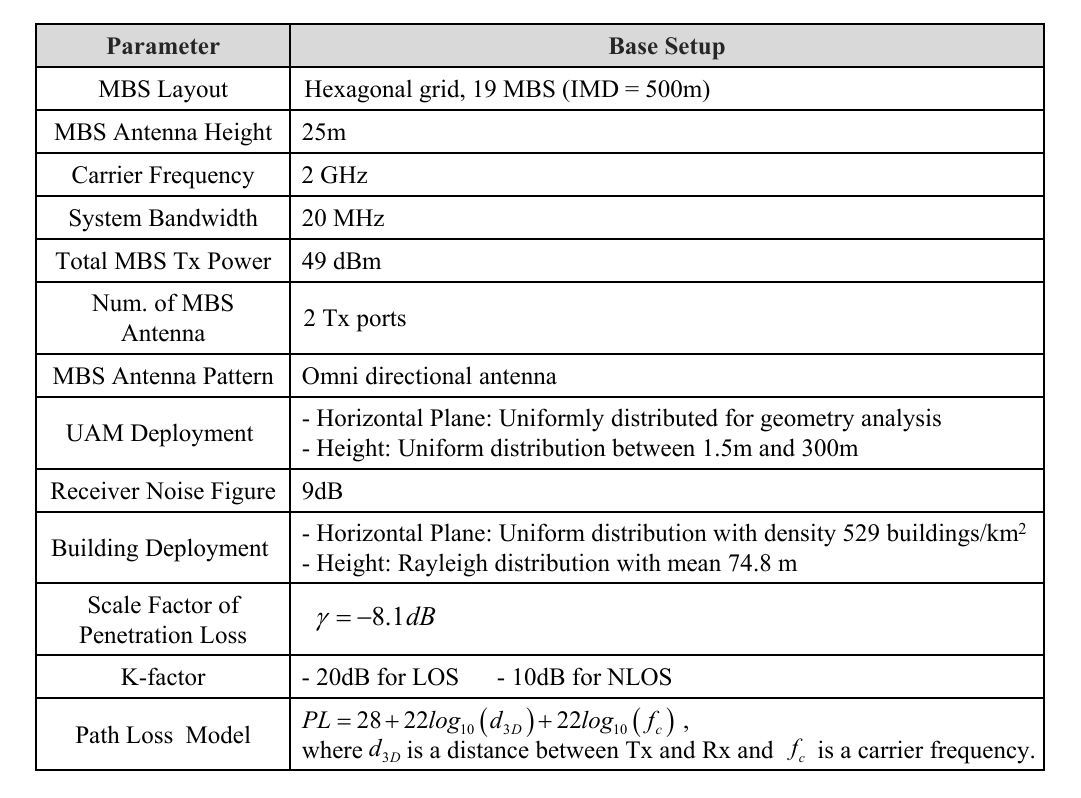}
\end{center}
\end{table}

\clearpage

\subsection{Performance Analysis for UAM Deployment} \label{SubSec_3.3.1}

In order to measure a geometry SINR distribution, we evenly deploy UAMs at 50 $m$ intervals over the entire area. 

\begin{figure}[h!]
    \centering
    \includegraphics[width=0.9\columnwidth]{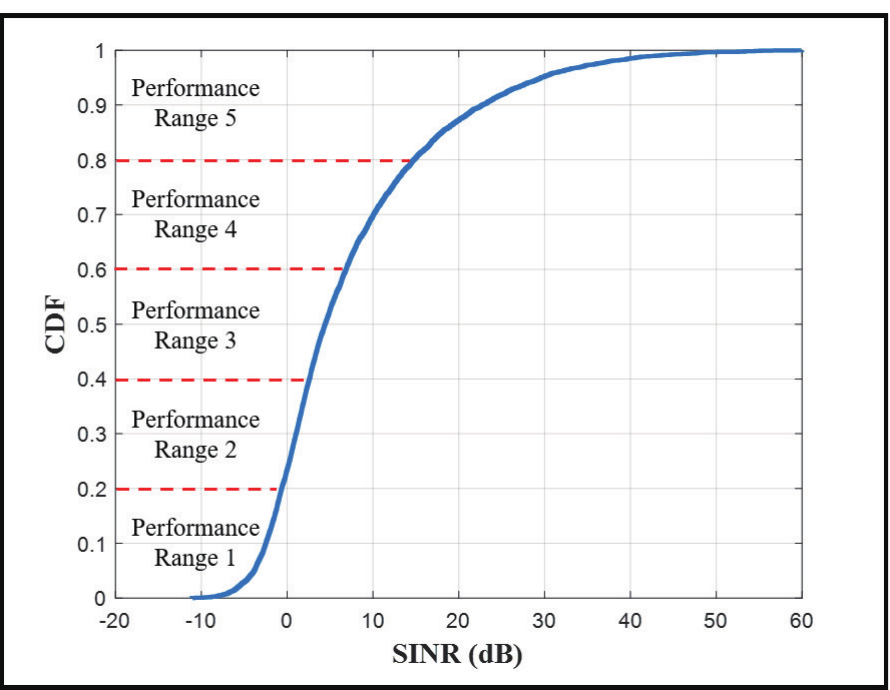}
    \caption{The SINR distribution of the received signal in a UAM network}
    \label{Fig3_3}
\end{figure}

As shown in Figure \ref{Fig3_3}, the SINR at a UAM has a distribution from $-10$ $dB$ to $60$ $dB$. In particular, it can be seen that the ratio of the UAMs below 0 dB SINR that can critically affect the reliability of the UAM network is about 20 $\%$. For a more detailed analysis in Figure \ref{Fig3_4} that follows, let's classify five performance ranges in units of $20$ $\%$. 

\clearpage

Figure \ref{Fig3_4} (a) depicts the average height of UAM for each performance range. In the lower performance range, the average height of the UAMs is higher. There are two reasons for this trend: i) the positive effect of blockages, and ii) being \textquotedblleft cell-edge" at high altitudes.

\begin{itemize}
    \item \textbf{The Positive Effect of Blockages}
    
    \,\,\,\,Figure \ref{Fig3_4} (b) shows the average number of buildings to block the signal transmitted from the serving MBS and interfering MBSs, respectively. The average number of blockages experienced by the desired signal is less than $1$, which is not a meaningful trend for each performance range. On the contrary, the average number of buildings blocking the interference signal is found to be higher in the case of UAM located in the middle of the building forest. In particular, in the case of the range $5$, the interfering signals pass through about $9.7$ buildings. As shown in the analysis of the horizontal condition of blockage events in \eqref{Average_N}, interfering MBS with a relatively far distance than serving MBS are more likely to experience blockage events. In addition, according to \eqref{total_Vertical_P}, as the altitude of a UAM increases, a UAM deviates from the blockage area; as a result, a UAM is exposed from the interfering signals. This shows that the buildings play a positive role in blocking interference signals.
    
     \item \textbf{Being \textquotedblleft Cell Edge" at High Altitude}
     
     \,\,\,\, In a conventional 2D cellular network, according to the horizontal movement of the user, it is close to a specific MBS and away from other MBSs. In addition, when the strength of the signal from the interfering MBSs is superior to the signal strength from the serving MBS, the region is called a cell edge. 
     
     \begin{figure}[p!]
    \centering
    \includegraphics[width=0.7\columnwidth]{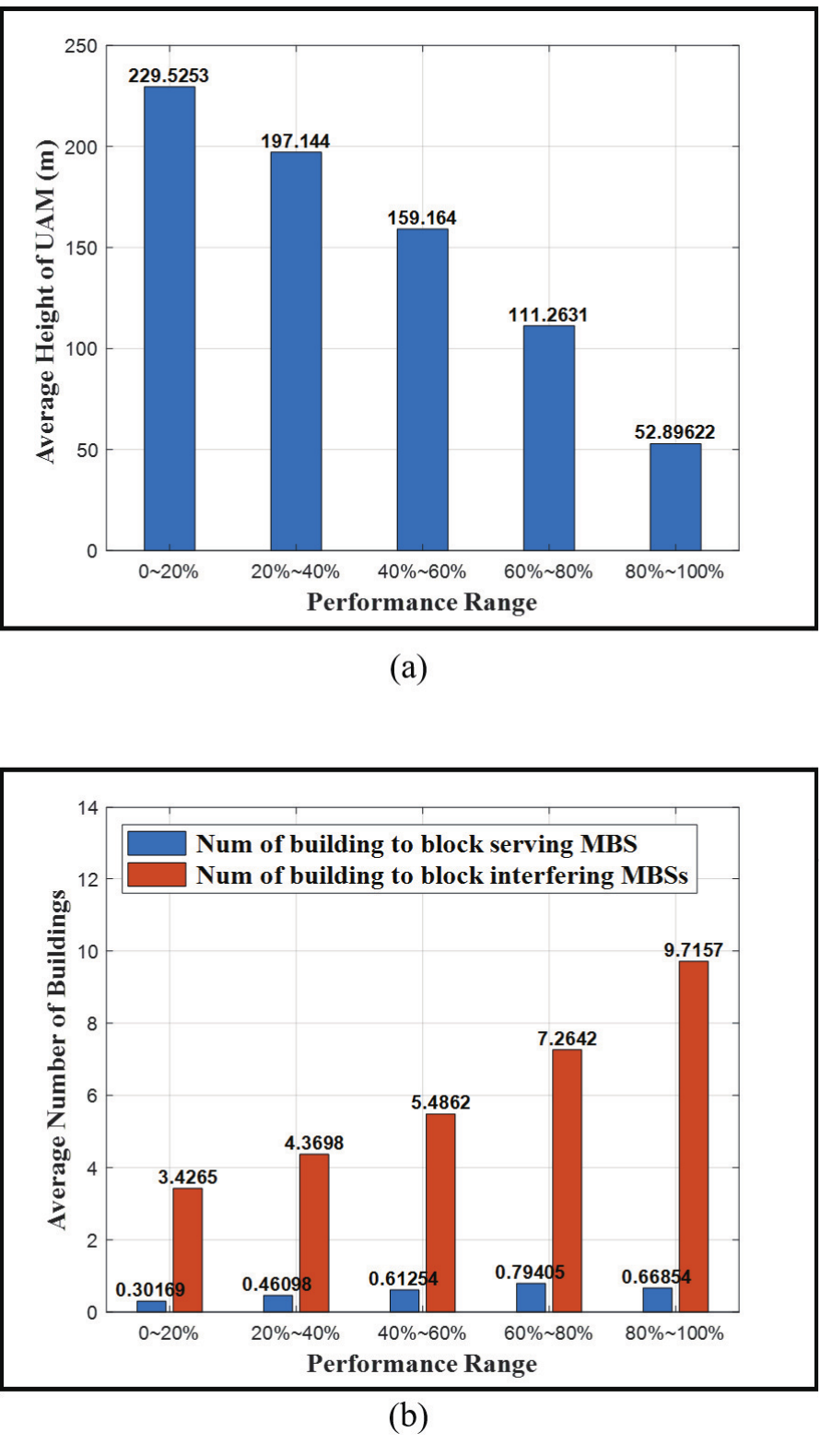}
    \caption{(a) the average height of UAM, (b) the average number of blocking buildings according to performance ranges.}
    \label{Fig3_4}
    \end{figure}
     
     \clearpage
     
     \,\,\,\, However, UAM systems with vertical mobility show a peculiar trend. When UAM rises, it moves away from serving MBS and interfering MBSs at the same time. As a result, above a certain altitude, there exists a region where the signal strength of the desired signal and the interference signal are the same. That is, at a high altitude, UAM recognizes it as a cell-edge environment.
     
     \,\,\,\, This phenomenon can be seen in Figure \ref{Fig3_5} (a), which is a visual representation of a coverage probability with the target SINR, $T=0$ $dB$. The green points indicate the positions of the UAM that satisfies $T$, and the blue points describe the UAMs that cannot. It can be intuitively seen that the UAMs with low SINR increases as the height of the UAM increases.

\end{itemize}

For these complex reasons, the coverage probability decreases depending on the UAM height, as shown in Figure \ref{Fig3_5} (b). In order to achieve high reliability, it is essential to improve the quality of high-altitude UAMs' signal above the average building height.

\begin{figure}[p!]
    \centering
    \includegraphics[width=0.7\columnwidth]{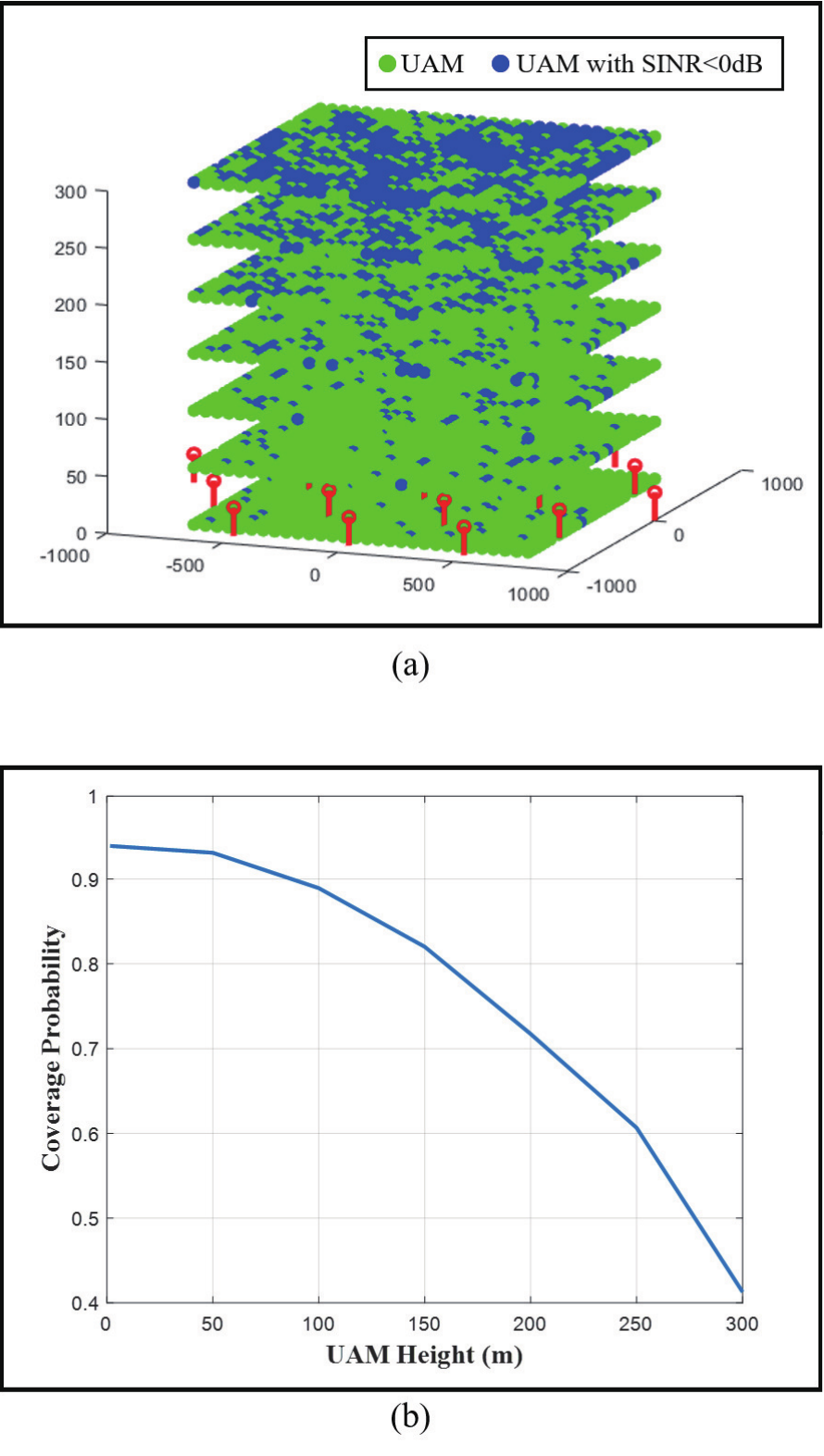}
    \caption{Coverage probability with the target SINR $T=0$ $dB$. (a) 3D visual representation, (b) coverage probability with various UAM height.}
    \label{Fig3_5}
\end{figure}

\clearpage
\subsection{Performance Analysis for Building Deployment} \label{SubSec_3.3.2}

The building is an environmental factor, not a system parameter we can control. However, in this subsection, we look at how buildings affect network performance, and consider how to safely operate UAM in a given environment.

\begin{figure}[p!]
    \centering
    \includegraphics[width=0.7\columnwidth]{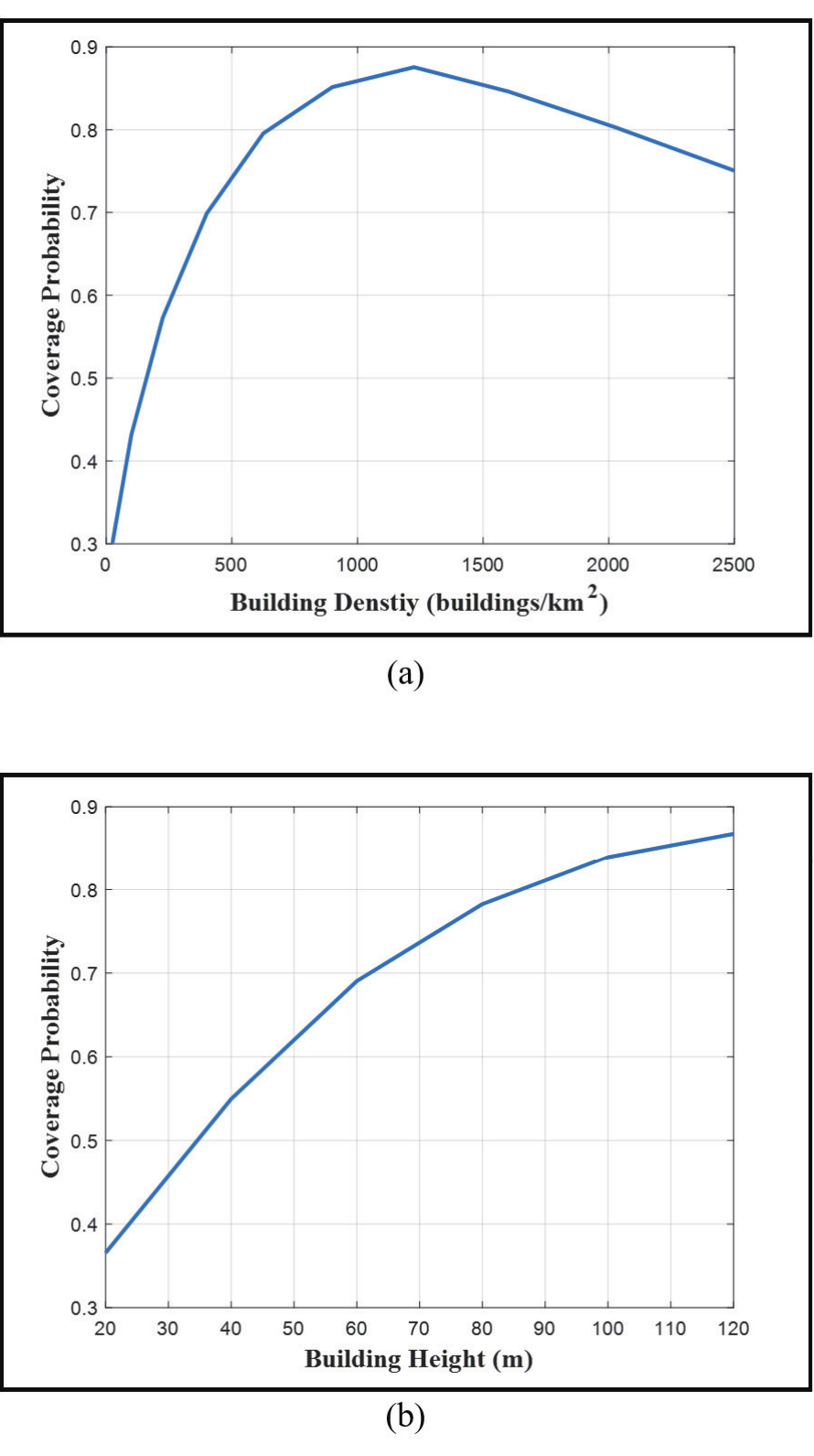}
    \caption{Coverage probability with various buildings' parameters: (a) the densities and (b) the heights.}
    \label{Fig3_6}
\end{figure}

\begin{itemize}

    \item \textbf{The Density of Buildings}
    
    \,\,\,\,Figure \ref{Fig3_6} (a) shows the coverage probability performance given building density. According to \eqref{PL_final}, the blockage probability linearly increases as the density of the buildings. The increase in the block probability of the interfering MBS signal is much steeper than the increase in the block probability of the serving MBS signal with a relatively close distance. As a result, the coverage probability performance increases due to the positive effect of the blockage, but when the point at which the desired signal is also blocked, the UAM of the lower performance rapidly decreases. Therefore, when considering the take-off and landing design of UAM, it is necessary to fully consider the density of the surrounding buildings.

    \item \textbf{The Height of Buildings}
    
    \,\,\,\, As shown in Figure \ref{Fig3_6} (b), a high-altitude building can play a positive role by blocking interference, but it is a very dangerous matter in relation to the safe operation of UAM. Therefore, only vertical take-off and landing should be allowed below the average height of the building, and horizontal movement should be made at a higher place than the height of the building. Therefore, UAM will still belong to the cell-edge, which is a challenge to overcome in designing UAM networks.

\end{itemize}

\clearpage
\subsection{Performance Analysis for MBS Deployment} \label{SubSec_3.3.3}

\begin{figure}[p!]
    \centering
    \includegraphics[width=0.7\columnwidth]{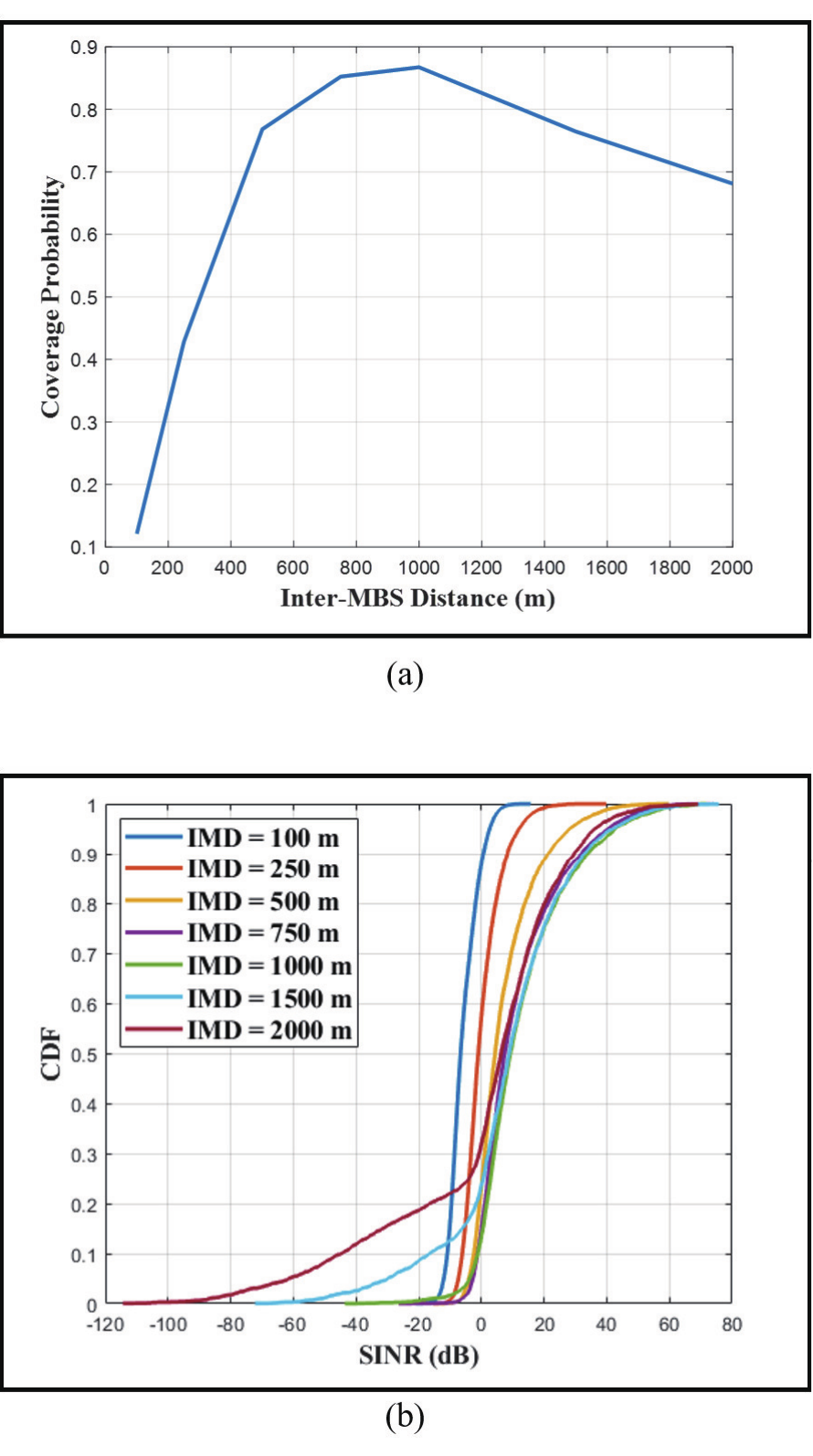}
    \caption{Performance with various IMDs. (a) coverage probability, (b) the CDF of SINR.}
    \label{Fig3_7}
\end{figure}

\begin{figure}[p!]
    \centering
    \includegraphics[width=0.7\columnwidth]{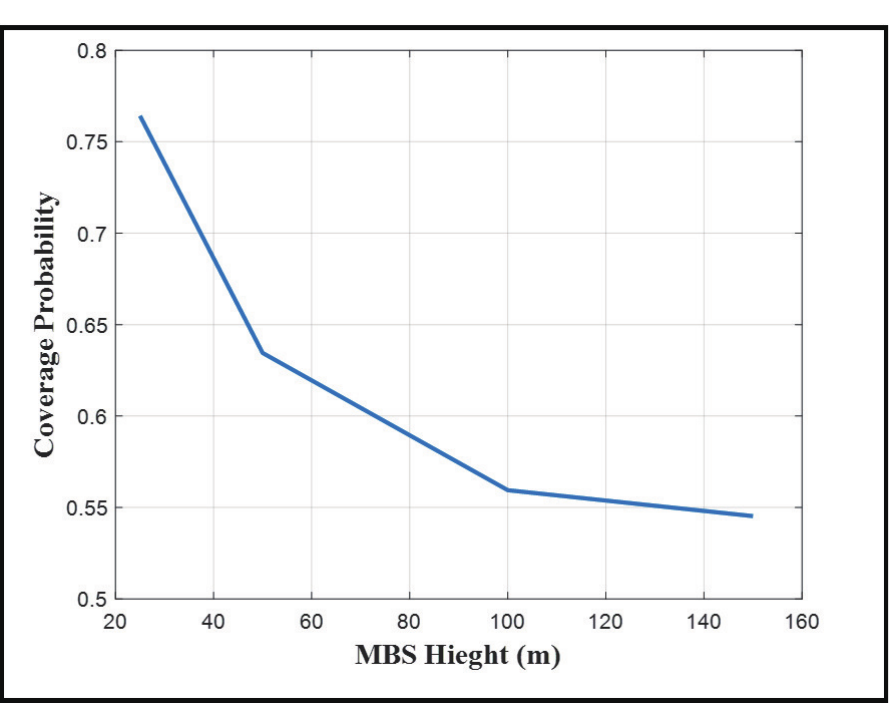}
    \caption{Coverage probability with various MBS height.}
    \label{Fig3_8}
\end{figure}

\begin{itemize}

    \item \textbf{Inter-MBS Distance}
    
    \,\,\,\,A simple method of reducing the cell-edge area at high altitudes is to increase the inter-MBS distance to reduce the effect of vertical mobility on SINR and make the horizontal distance more dominant. Figure \ref{Fig3_7} shows the coverage probability performance in various IMDs. In terms of coverage performance, the best performance is shown when the IMD is $1$ $km$. After $1$ $km$, it can be seen that the SINR drops sharply due to the decrease in the receiving power. 
    
    \,\,\,\,It is important, however, to maximize the minimum SINR as well as coverage probability in the UAM network of which reliability is the most important KPI. Therefore, given the building deployment in Gang-Nam Station, it can be said that $750$ $m$ of IMD is suitable for MBS deployment.
    
    \item \textbf{The Height of MBS}
    
    \,\,\,\,Figure \ref{Fig3_8} the coverage probability with various MBS height. As the height of the MBS increases, the height difference from the UAM decreases as it goes to the top of the building. This causes an increase in the height of the penetration point in the vertical condition \eqref{total_Vertical_P} and decreases the benefit of blockage events. Therefore, in order to make the most of the blockage phenomenon, the lower height of MBSs leads to better network performance.

\end{itemize}

\clearpage

\clearpage
  



\chapter{Link Reliability Improvement for Urban Air Mobility Communications}\label{Chap_4}
\hspace{-5mm}\rule{135mm}{1pt}

{4.1 The Optimal Elevation Tilting of the Antenna 

4.2 Frequency Reuse with Multi-layered Narrow Beam  

4.3 The Assistive Transmission of a Master UAM for Achieving 99.9 \% Reliability\\}
\rule{135mm}{1pt}
\clearpage

\section{The Optimal Elevation Tilting of the Antenna}\label{Sec_4.1}

The Omni-directional antennas, in Chapter 3, are considered to analyze the urban environment and to find out the basic performance of UAM networks. In this chapter, we adopt a directional antenna to reduce interference to neighboring MBSs while concentrating the energy of the desired signal to the target UAM, which is susceptible to interference at a high altitude.

The following is an equation explaining the pattern of the directional antenna gain defined in the 3GPP standard \cite{27}.
\begin{equation}\label{AG}
\begin{array}{*{20}{c}}
{AG({\varphi _{UM}},\,\,\varphi '\,,\,\,{\theta _{UM}}) =  - min\left\{ { - \left[ {A{G_V}\left( {{\varphi _{UM}},\,\,\varphi '} \right) + A{G_H}\left( {{\theta _{UM}}} \right)} \right],\,\,{A_m}} \right\},}\\
\begin{array}{l}
{\rm{where}}\,\,\,A{G_V}\left( {{\varphi _{UM}},\,\,\varphi '} \right) =  - min\left\{ {12\left( {\frac{{{\varphi _{UM}} - \varphi '}}{{{\varphi _{3dB}}}}} \right)\,,\,\,SLA_{V}} \right\}\,,\,\\
\,\,\,\,\,\,\,\,\,\,\,\,\,\,\,\,\,\,\,A{G_H}\left( {{\theta _{UM}}} \right) =  - min\left\{ {12\left( {\frac{{{\theta _{UM}}}}{{{\theta _{3dB}}}}} \right)\,,\,\,{A_m}} \right\}\,,\\
\,\,\,\,\,\,\,\,\,\,\,\,\,\,\,\,\,\,\,{\varphi _{3dB}} = 65^\circ ,\,\,SLA_{V} = 30dB,\,\,{\theta _{3dB}} = 65^\circ ,\,\,{A_m} = 30dB,
\end{array}
\end{array}
\end{equation}
$AG_V$ is a vertical pattern of MBS antenna and $AG_H$ is a horizontal pattern of MBS antenna. ${\varphi _{UM}} = {\tan ^{ - 1}}\left( {{{{{{h_M} - {h_U}} \mathord{\left/
 {\vphantom {{{h_M} - {h_U}} r}} \right.
 \kern-\nulldelimiterspace} r}}_{UM}}} \right)$ is the vertical angle of the line UAM-MBS $\overline {UM}$, and  ${\theta _{UM}} = {\tan ^{ - 1}}\left( {{{\left( {{y_U} - {y_M}} \right)} \mathord{\left/
 {\vphantom {{\left( {{y_U} - {y_M}} \right)} {\left( {{x_U} - {x_M}} \right)}}} \right.
 \kern-\nulldelimiterspace} {\left( {{x_U} - {x_M}} \right)}}} \right)$ is the horizontal angle of one. $\varphi _{3dB}$ and $\theta _{3dB}$ represent half of the vertical and horizontal beam-widths, respectively. $SLA_{V}$ and $A_m$ stand for the vertical and horizontal side-lobe suppression in dB, respectively. $\varphi '$ refers to the elevation tilting angle of the antennas.

\begin{figure}[p!]
    \centering
    \includegraphics[width=0.7\columnwidth]{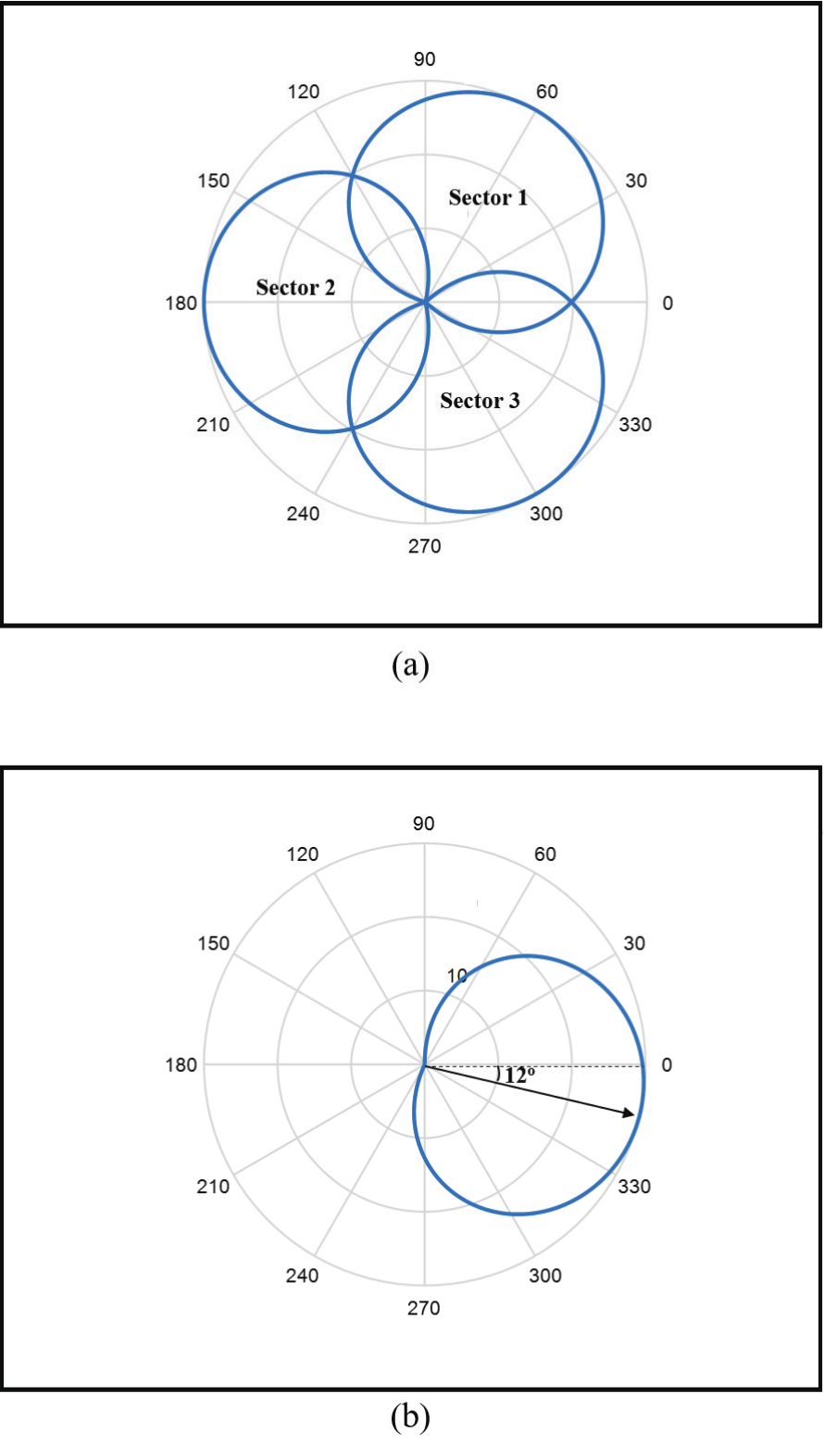}
    \caption{The 3GPP antenna gain pattern: (a) 3 sector in horizontal plane, (b) vertical plane.}
    \label{Fig4_1}
\end{figure}

Figure \ref{Fig4_1} depicts the antenna setup for the conventional terrestrial communication system currently defined by 3GPP standard. As shown in (a), it divides into three sectors, concentrating the propagation to each target area, and maximizing frequency reuse. In addition, the angle of the antenna is optimized to 12 degrees downward for ground users. 

However, this setup is not suitable for serving high-level UAMs. It is necessary to analyze the tilting angle of the antenna according to each altitude. By maximizing the sum of the antenna gains for the service area, the optimal tilting angle for a given height can be obtained. Given $h_U$, the optimal elevation tilting angle $\varphi '$ can be derived as

\begin{equation}\label{Optimal_angle}
[\varphi '\left| {{h_U}} \right.] = \mathop {arg\,max}\limits_{\varphi '\,} \int\limits_{{r_{UM}} = 0}^{IMD/2} {\int\limits_{{\theta _{UM}} =  - {\theta _{3dB}}}^{{\theta _{3dB}}} {AG({\varphi _{UM}},\,\,\varphi '\,,\,\,{\theta _{UM}}\left| {{h_U}} \right.)d{\theta _{UM}}d{r_{UM}}} } .
\end{equation}

\begin{figure}[b!]
    \centering
    \includegraphics[width=0.7\columnwidth]{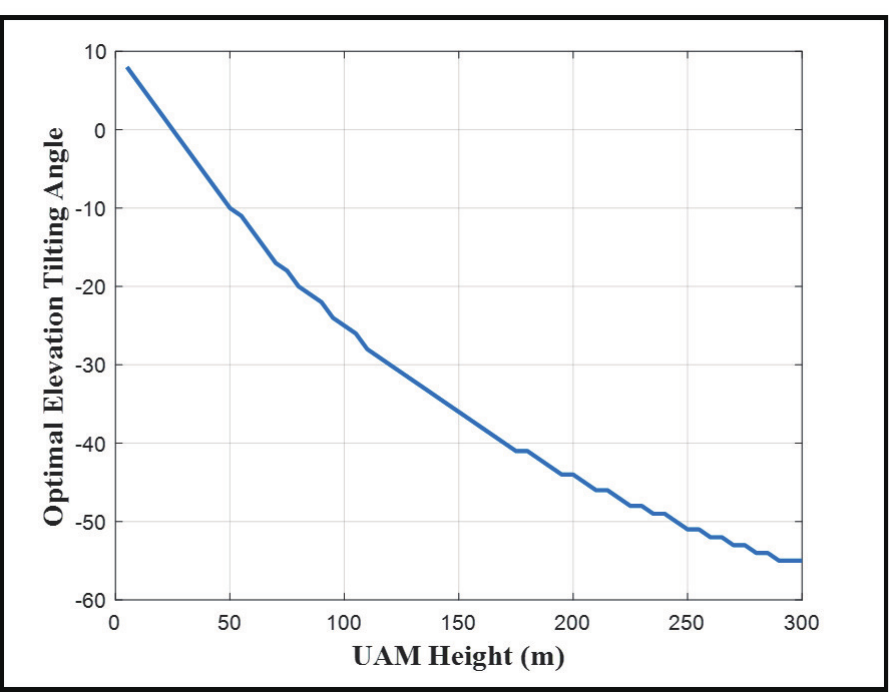}
    \caption{The optimal tilting angle for a given UAM height.}
    \label{Fig4_2}
\end{figure}

Figure \ref{Fig4_2} shows the optimal tilting angle for the given UAM height. A positive angle means a downward angle, and a negative number means an upward angle. As the elevation of the UAM increases, the angle to be targeted also increases so that the tilting angle gradually decreases.

\clearpage

\begin{figure}[b!]
    \centering
    \includegraphics[width=0.7\columnwidth]{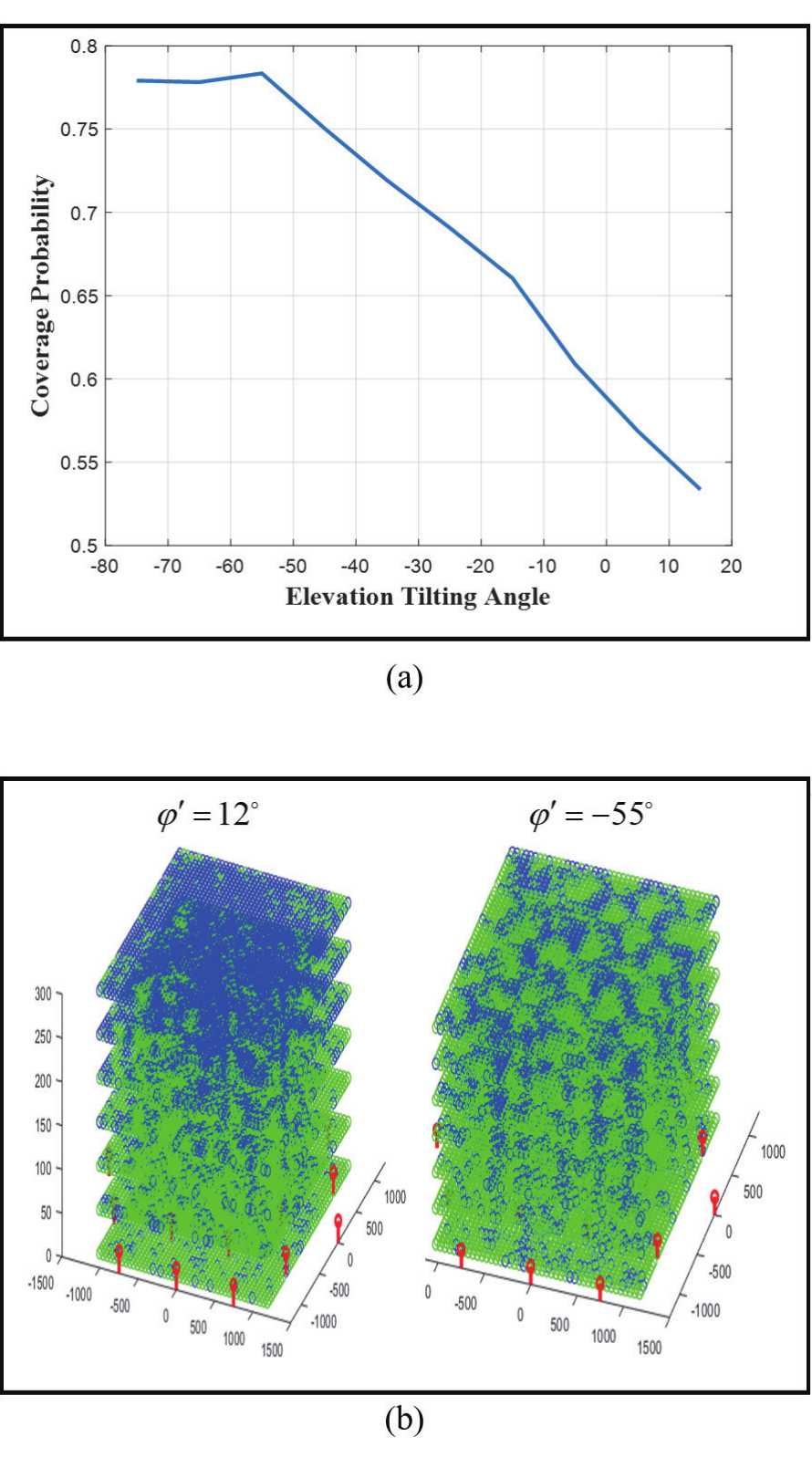}
    \caption{(a) the coverage probability performance with a given tilting angle and (b) a visualized presentation}
    \label{Fig4_3}
\end{figure}

The main goal of UAM networks is not to improve the overall average performance, but to maximize reliability. Therefore, we have to optimize the antenna angle to focus on the UAMs with lower performance. Figure \ref{Fig4_3} (a) verify the coverage probability performance with a given tilting angle. It is demonstrated that the antenna setting targeting the lowest performance UAM located at the highest altitude shows the best effect on the coverage performance. Figure \ref{Fig4_3} (b) is a visual representation of the result of (a), and it can be confirmed that the high-altitude UAM performance is significantly improved. However, the performance degradation due to the cell edge area at high altitudes still exists, indicating that it is insufficient to satisfy the UAM reliability requirements.

\clearpage
\section{Frequency Reuse with Multi-layered Narrow Beam} \label{Sec_4.2}

\begin{figure}[b!]
    \centering
    \includegraphics[width=0.6\columnwidth]{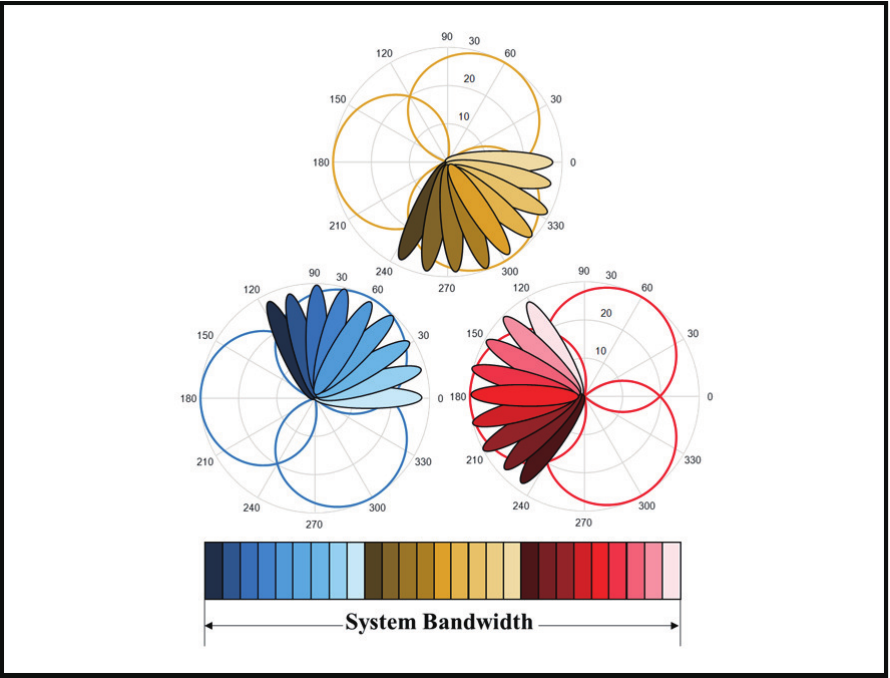}
    \caption{The Antenna Gain Pattern with frequency reuse factor $\delta=3$}
    \label{Fig4_4}
\end{figure}

In the previous section, we can see that despite adopting the optimal inclination angle, it is still impossible to overcome the cell edge region. To handle with cell edge, as shown in Figure \ref{Fig4_4}, we propose a frequency reuse pattern with narrow beamforming. To be specific, it is designed to dramatically reduce interference between adjacent MBSs through beam avoidance and resource splitting. Given the number of beams per sector $N_{beam}$, the horizontal gain of the $n_{th}$ beam can be derived from \eqref{AG} as

\begin{equation}\label{Beam}
\begin{array}{*{20}{c}}
{A{G_H}\left( {{\theta _{UM}},\,\,n} \right) =  - min\left\{ {12\left( {\frac{{{\theta _{UM}} + 60^\circ  + {\theta _{3dB}} - 2n{\theta _{3dB}}}}{{1.083\, \times {\theta _{3dB}}}}} \right)\,,\,\,{A_m}} \right\}\,,}\\
{{\rm{where}}\,\,\,{\theta _{3dB}} = \frac{{60^\circ }}{{{N_{beam}}}},\,\,{A_m} = 30dB}.
\end{array}
\end{equation}
From the strongest RSRP beam in \eqref{AG} and \eqref{Beam}, The transmit signal $X$ is written by
\begin{equation}\label{X}
X = \mathop {max}\limits_{n \in {N_{beam}}} \left[ {AG({\varphi _{UM}},\,\,\varphi '\,,\,\,{\theta _{UM}},\,\,n)} \right]x,
\end{equation}
where $x$ is original transmit data.

\clearpage

\begin{itemize}
    \item \textbf{Coverage Probability Analysis}
    
    \,\,\,\,For tractable analysis, we assume that the frequency reuse pattern is randomly chosen. Given frequency reuse factor $\delta$, the cumulative interference can be approximated as decreasing by $\delta{N_{beam}}$ times. As a special case ${N_{beam}}=1$, the UAM network obtains performance gain only through frequency reuse without beamforming. Based on \eqref{Coverage_result} and \eqref{Laplace}, the coverage probability with frequency reuse can be represented as 
    \begin{equation}\label{Coverage_FR}
    \begin{array}{l}
    {P_c}(T,\,\delta ,{N_{beam}}) = \int\limits_{{h_U} = {h_U^{\min }}}^{{h_U^{\max }}} {\int\limits_{{r_{UM}} = 0}^\infty  {\sum\limits_{p = 0}^{m - 1} {\frac{{{{\left( { - sAG({\varphi _{UM}},\,\,\varphi '\,,\,\,{\theta _{UM}})} \right)}^p}}}{{p!}}} } } \\
    \,\,\, \times \left[ {\frac{{{\partial ^p}}}{{\partial {s^p}}}{L_{{I_{\delta {N_{beam}}}}}}\left( {s\left| {{r_{UM}},\,} \right.\,{h_U}} \right)} \right]{f_{{r_{UM}}}}({r_{UM}}){f_{{h_U}}}({h_U})d{r_{UM}}d{h_U},
    \end{array}
    \end{equation}
    where
    \begin{equation}\label{Lapalce_FR}
    \begin{array}{l}
    {L_{{I_{\delta {N_{beam}}}}}}\left( {s\left| {{r_{UM}},\,{h_U}} \right.} \right) = \exp \left[ { - 2\pi \frac{{{\lambda _M}}}{{\delta {N_{beam}}}}\int\limits_{{r_{UM}}}^\infty  {{e^{{ - ^{E\left[ {{N_{BP}}\left( {{r_{UM}},\,{h_U}} \right)} \right]}}}}} } \right.\\
    \,\,\,\,\,\, \times \left. {\left[ {1 - {{\left( {1 + \frac{{sAG(\varphi _{UM}^{ji},\,\,\varphi '\,,\,\,\theta _{UM}^{ji}){{\left( {r_{UM}^{ji}} \right)}^{ - \frac{\alpha }{2}}}}}{m}} \right)}^{ - m}}} \right]tdt} \right]
    \end{array}.
    \end{equation}
 
     \,\,\,\,Figure \ref{Fig4_5} shows coverage probability with $\delta=3$ and various number of beams. As shown in \eqref{Coverage_FR}, when ${N_{beam}}$ increases, the beam narrows, improving the interference avoidance effect, and at the same time, the number of interfering MBSs decreases because resources are divided and used for each beam. The key observation being that more than 3 beams already provide sufficient interference avoidance, the performance of the coverage probability is saturated.
    
    \begin{figure}[p!]
    \centering
    \includegraphics[width=0.9\columnwidth]{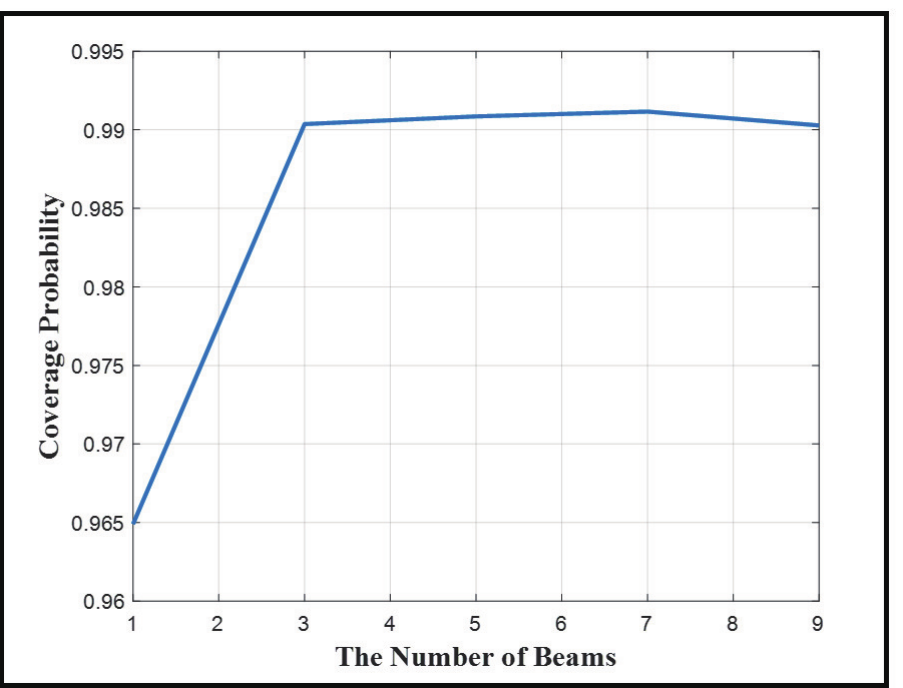}
    \caption{The coverage probability with $\delta=3$ and various number of beams.}
    \label{Fig4_5}
\end{figure}

\clearpage

    \item \textbf{Reliability Analysis}
    
    \,\,\,\, Reliability is defined as the probability of successful transmission within the required latency at the application layer while under network coverage. Given latency constraint $t_R$ and the C2 packet size $\tau_R$, the reliability $R$ can be expressed by
    
    \begin{equation}\label{Reliability_definition}
    R = P\left( {\tau  > {\tau _R}\left| {{t_R}} \right.} \right)
    \end{equation}
    where $\tau$ is achievable rate as following,
    \begin{equation}\label{Rate}
    \begin{array}{l}
    \tau  = \frac{{BW}}{{\delta {N_{beam}}N_U^{\sec tor}}}E\left[ {\ln \left( {1 + SIR} \right)} \right]\\
    \,\,\,\, = \frac{{BW}}{{\delta {N_{beam}}N_U^{\sec tor}}}\int\limits_{{h_U} = h_U^{\min }}^{h_U^{\max }} {\int\limits_{{r_{UM}} = 0}^\infty  {\sum\limits_{p = 0}^{m - 1} {\frac{{{{\left( { - zAG({\varphi _{UM}},\,\,\varphi '\,,\,\,{\theta _{UM}})} \right)}^p}}}{{p!}}} } } \\
    \,\,\,\,\,\,\,\,\, \times \left[ {\frac{{{\partial ^p}}}{{\partial {z^p}}}{L_{{I_{\delta {N_{beam}}}}}}\left( {z\left| {{r_{UM}},\,} \right.\,{h_U}} \right)} \right]{f_{{r_{UM}}}}({r_{UM}}){f_{{h_U}}}({h_U})d{r_{UM}}d{h_U}
    \end{array}
    \end{equation}
    denoting ${z = {P_{TX}}{L_{UM}}{{\left( {{{\left( {{r_{UM}}} \right)}^2} + {{\left( {{h_U} - {h_M}} \right)}^2}} \right)}^{{\raise0.7ex\hbox{$\alpha $} \!\mathord{\left/
     {\vphantom {\alpha  2}}\right.\kern-\nulldelimiterspace}
    \!\lower0.7ex\hbox{$2$}}}}\left( {{e^t} - 1} \right)}$. $BW$ and ${N_U^{\sec tor}}$ is a system bandwidth and the number of UAMs per a MBS sector, respectively.
    
    \,\,\,\,For a realistic reliability evaluation than the geometry-based performance analysis previously performed, let's consider a simulator that reflects the 3D mobility of UAMs. As illustrated in Figure \ref{Fig4_6}, the take-off points and landing points of UAMs are defined 2D PPP model as ${\Phi _{{U_T}}} = \left\{ {{U_{{0_T}}},\,{U_{{1_T}}},\, \cdots ,\,{U_{{{\left( {{N_U} - 1} \right)}_T}}}} \right\}$ and ${\Phi _{{U_L}}} = \left\{ {{U_{{0_L}}},\,{U_{{1_L}}},\, \cdots ,\,{U_{{{\left( {{N_U} - 1} \right)}_L}}}} \right\}$, respectively. The maximum height of the $j_{th}$ UAM has random uniform distribution between $100$ and $300$ $m$. The vertical and horizontal speeds of UAMs are set to $50$ and $160$ $km/h$, respectively. In addition, the number of UAMs that can be operated simultaneously is limited to 50 units.

    \begin{figure}[p!]
        \centering
        \includegraphics[width=0.7\columnwidth]{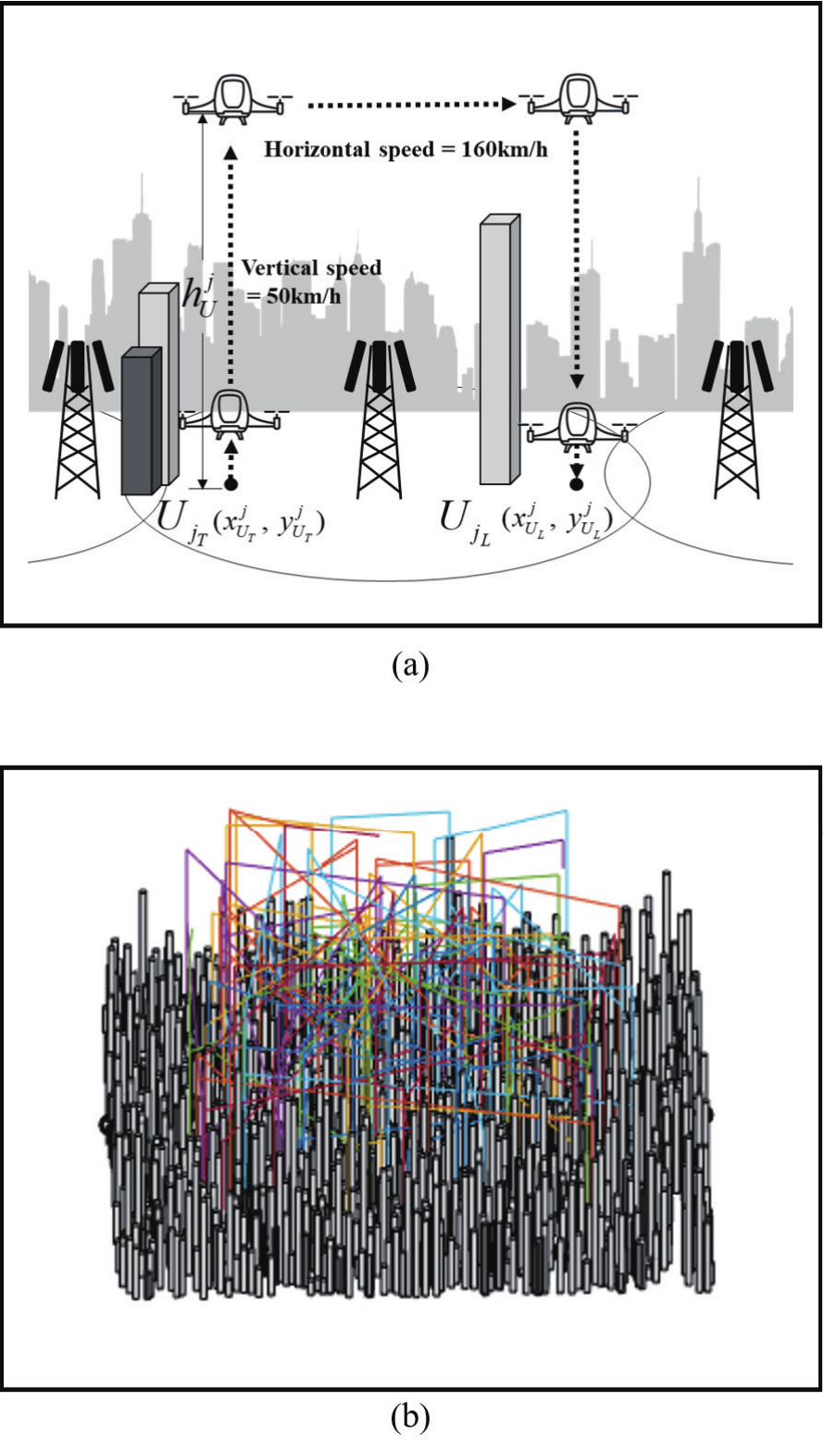}
        \caption{(a) The PPP modeling for 3D mobility of UAMs, (b) the example of simulation with 3D mobility.}
        \label{Fig4_6}
    \end{figure}
    
    \clearpage
    
    \,\,\,\,Figure \ref{Fig4_7} shows reliability performance with various number of beams per sector. It can be seen that the more the beams are sharpened, the better SINR can be obtained due to the desired signal consecration and interference avoidance. In the horizontal movement of UAM, the effect of beam splitting is prominent. However, since resources are divided as many as the number of beams, it can be seen that more than three beams per sector provide inefficient reliability performance. In particular, as shown in Table \ref{Table1_1}, since the reliability required for infrastructure access is very high, a decrease in the bandwidth that can be used by individual beams causes performance degradation.
    
    \begin{figure}[h!]
        \centering
        \includegraphics[width=0.7\columnwidth]{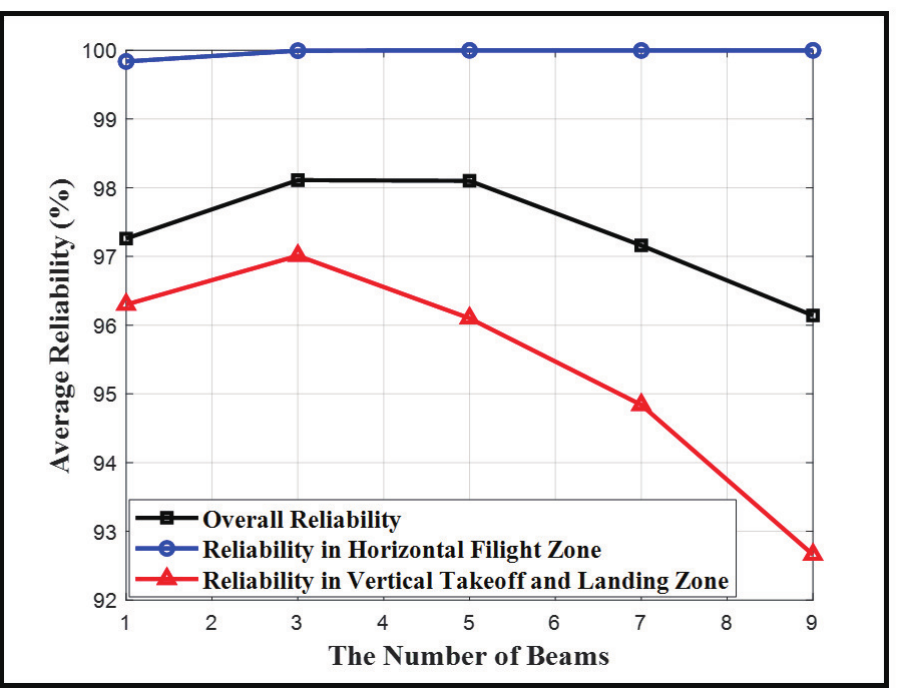}
        \caption{The reliability performance with various number of beams. The black line denotes the overall reliability performance, and the blue and red lines describes reliability in horizontal and vertical movements, respectively.}
        \label{Fig4_7}
    \end{figure}
    
    \clearpage
    
    \item \textbf{Proposed Multi-layered Antenna Gain and Resource Splitting Pattern}
    
    \,\,\,\, The UAM network is classified into a horizontal autonomous flight zone and a VTOL zone. In the autonomous flight zone with a strong line of sight characteristic, not only the desired signal but also the interference signal is strongly received, so the interference can be effectively controlled through the beam and resource splitting described before. Also, since it has low-requirement for automatic flight C2 messages reliability, the rate drop due to resource splitting is negligible. On the other hand, as shown in Figure \ref{Fig4_7}, resource splitting may be inefficient because the VTOL zone requires a high level of reliability. Fortunately, we saw in Chapter 3 that buildings in VTOL zones play a role in preventing interference. Through these observations, we propose a vertical antenna pattern composed of an upward beam for the horizontal zone and a downward beam for the VTOL zone as shown in Figure \ref{Fig4_8} (a). The tilting angle of each beam can be obtained through the equation \eqref{Optimal_angle}. Figure \ref{Fig4_8} (b) describes the final proposed multi-layer antenna structure.  The upward beam adopts the resource splitting pattern in Figure \ref{Fig4_4}, while the downward beam utilizes the whole bandwidth allocated for each sector’s corresponding.
    
    \,\,\,\, Figure \ref{Fig4_9} shows he performance of the proposed multi-layer antenna gain pattern. Although the coverage probability is 0.98, which is lower than the performance of single vertical layer with resource splitting of 0.99 in Figure \ref{Fig4_5}, it can be seen that the average reliability performance is improved to 99.65 \% by using the whole bandwidth for the downward beam.

    \begin{figure}[p!]
        \centering
        \includegraphics[width=0.65\columnwidth]{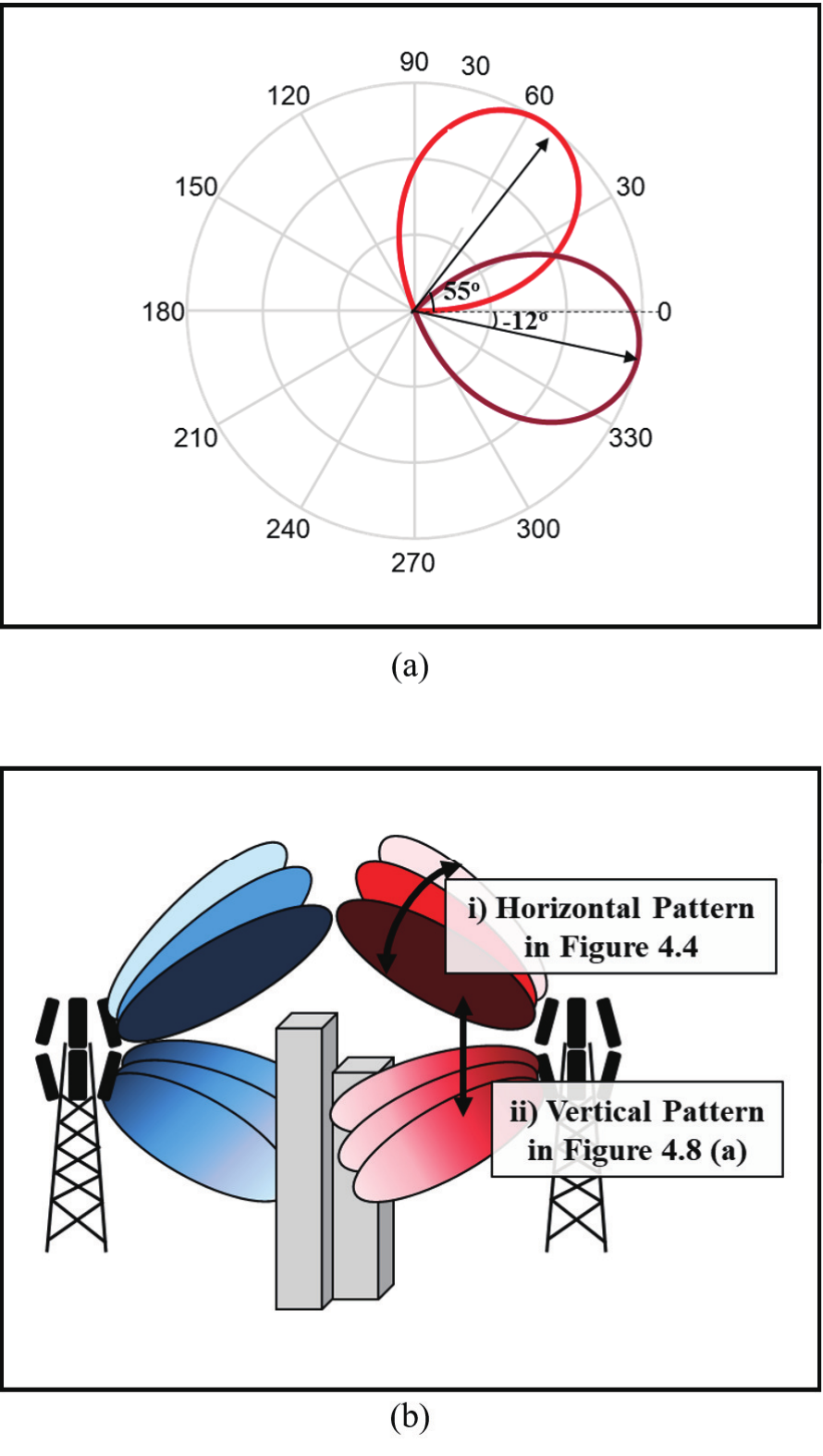}
        \caption{(a) vertical antenna pattern which consists of an upward beam for supporting horizontally moving UAM and a downward beam for vertically moving UAM. (b) the final proposed multi-layer antenna structure. The upward beam adopts the resource splitting pattern in Figure 4.4, while the downward beam utilizes whole bandwidth allocated for each sector's corresponding.}
        \label{Fig4_8}
    \end{figure}

      \begin{figure}[p!]
        \centering
        \includegraphics[width=0.7\columnwidth]{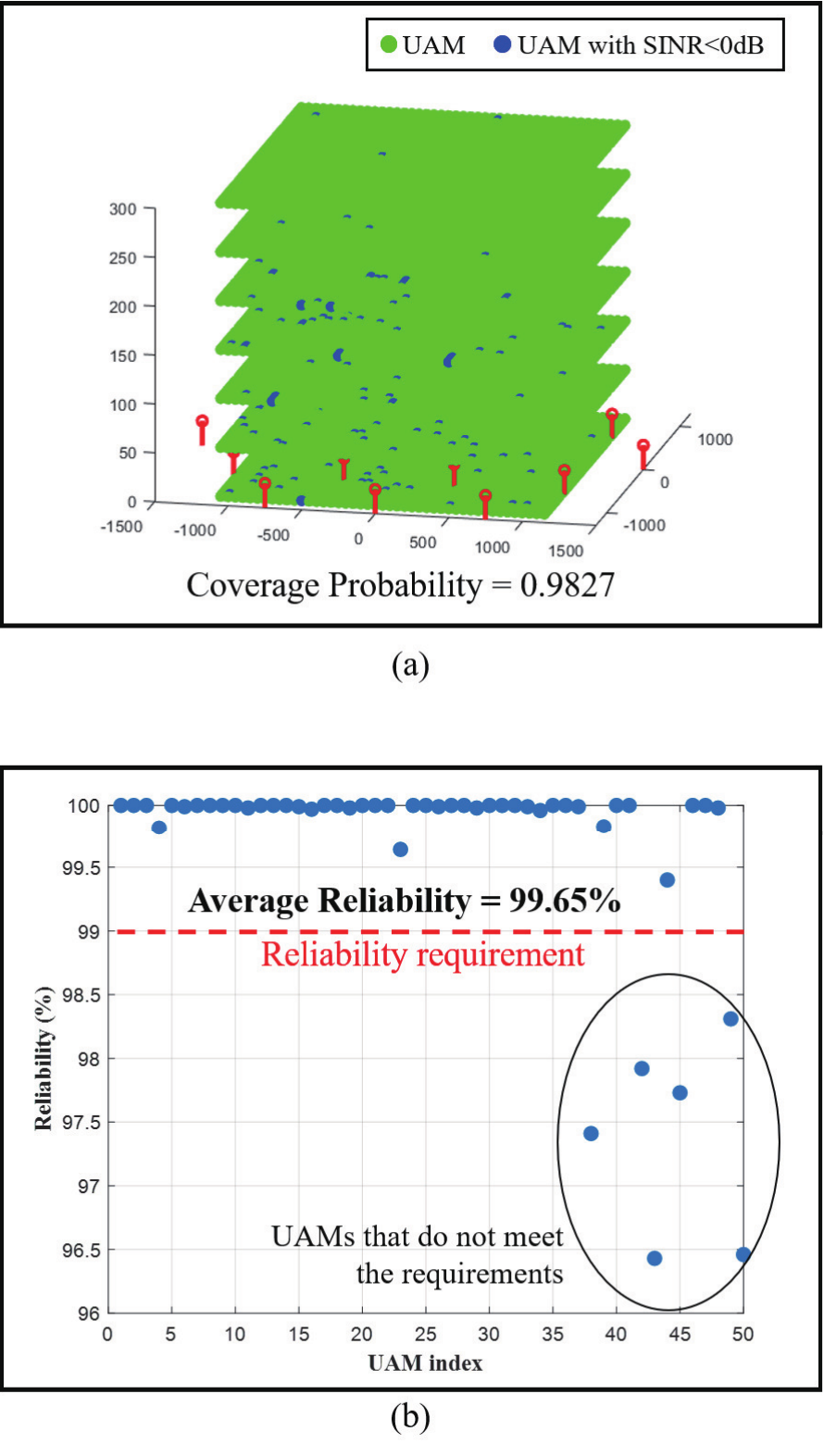}
        \caption{The performance of the proposed multi-layer antenna gain pattern: (a) visualized coverage probability, and (b) reliability performance of each UAM.}
        \label{Fig4_9}
    \end{figure}
    
    \clearpage
    
    \item \textbf{The Number of Supportable UAMs per Sector}
    
    \,\,\,\, From the equation \eqref{Coverage_main} and \eqref{Reliability_definition}, the number of supportable UAMs per sector $N_U^{\sec tor}$ can be approximated by 
    
        \begin{equation}\label{Supportable_UAMs}
    N_U^{\sec tor} = \frac{{{{{\tau _R}} \mathord{\left/
 {\vphantom {{{\tau _R}} {{t_R}}}} \right.
 \kern-\nulldelimiterspace} {{t_R}}}}}{{\frac{{BW}}{{\delta {N_{beam}}}}\ln \left( {1 + {P_c}^{ - 1}(R)} \right)}},
    \end{equation}
    
    where $T_R$ is the target SINR to satisfy the reliablity $R$ and ${P_c}^{ - 1}(R)$ is the inverse function of coverage probability in \eqref{Coverage_main}. Figure \ref{Fig4_10} shows the numerical results of the number of supportable UAMs corresponding to average reliability when the system bandwidth is 20 MHz and frequency reuse factor is 3. Under 99 \% reliability, about 1.7 UAM per sector can be supported.

          \begin{figure}[h!]
        \centering
        \includegraphics[width=0.7\columnwidth]{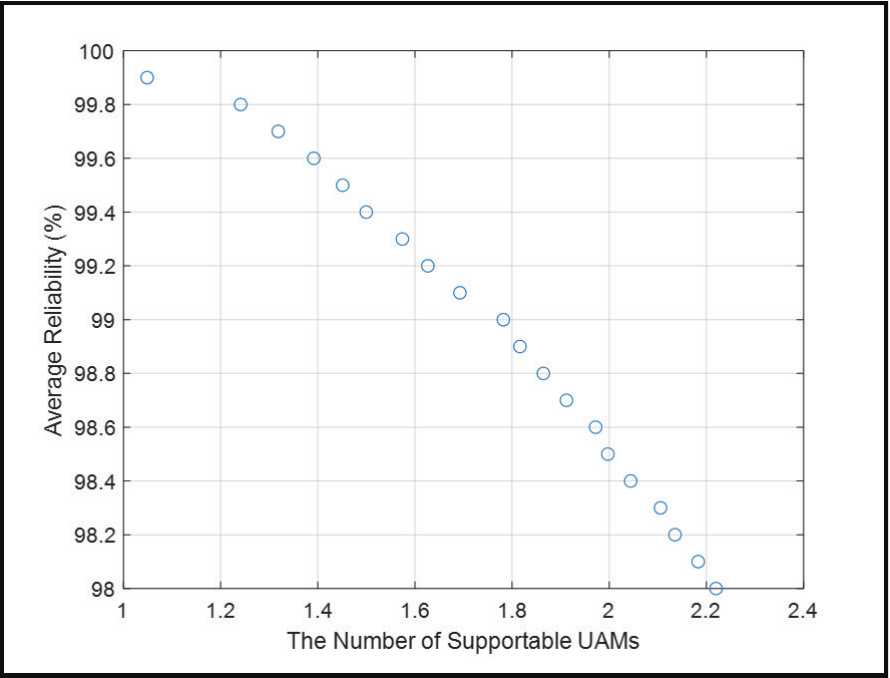}
        \caption{The Number of Supportable UAMs per Sector.}
        \label{Fig4_10}
    \end{figure}

\end{itemize}

\clearpage

\section{The Assistive Transmission of a Master UAM \\ for Achieving 99.9 \% Reliability}\label{Sec_4.3}

Even though we can achieve about 99 \% reliability via the optimal tilting and frequency reuse with multi-beams, it does not completely overcome the shadowing effect and cell edge in an urban environment. In this section, we propose an assistive transmission utilizing a master UAM to achieve the last 1 \% reliability. 

\begin{itemize}
\item \textbf{The Main Concept of the Master UAM Transmissions}

\,\,\,\,The MBS that can provide high quality links near UAMs is UAM itself. In the proposed protocol, as shown in Figure \ref{Fig4_8}, the UAM with the reserve of resources is defined as the mater UAM because the rate performance is higher than the target rate $\tau_R$, and this UAM receives and broadcasting the essential control signals for other UAMs from the serving MBS. In order to eliminate interference from MBSs, it is assumed that broadcasting uses different frequency from MBSs' one. 

\,\,\,\,Given the mater UAM $U_m$, the SNR of the $j_{th}$ UAM can be expressed by
\begin{equation}\label{UU_SNR}
{\rm{SNR(}}r_{UU}^{jm},\,\,h_U^j{\rm{,}}\,\,h_U^m{\rm{) = }}\frac{{{P_{T{X_U}}}G_{UU}^{jm}L_{UU}^{jm}{{\left( {{{\left( {r_{UU}^{jm}} \right)}^2} + {{\left( {h_U^j - h_U^m} \right)}^2}} \right)}^{{\raise0.7ex\hbox{${ - \alpha }$} \!\mathord{\left/
 {\vphantom {{ - \alpha } 2}}\right.\kern-\nulldelimiterspace}
\!\lower0.7ex\hbox{$2$}}}}}}{{{\sigma ^2}}},
\end{equation}
where ${{P_{T{X_U}}}}$ is transmit power of the master UAM.

\clearpage

\,\,\,\,Assuming equal scheduling, the data rate that the master UAM can additionally provide is as follows.
\begin{equation}\label{UU_Rate}
{\tau _{UU}} = \frac{1}{{{N_{{U_F}}}}}E\left[ {\ln \left( {1 + {\rm{SNR(}}r_{UU}^{jm},\,\,h_U^j{\rm{,}}\,\,h_U^m{\rm{)}}} \right)} \right],
\end{equation}
where $N_{U_F}$ is the number of failure UAMs not to success transmitting the target rate $\tau_R/t_R$.

\begin{figure}[h!]
    \centering
    \includegraphics[width=0.9\columnwidth]{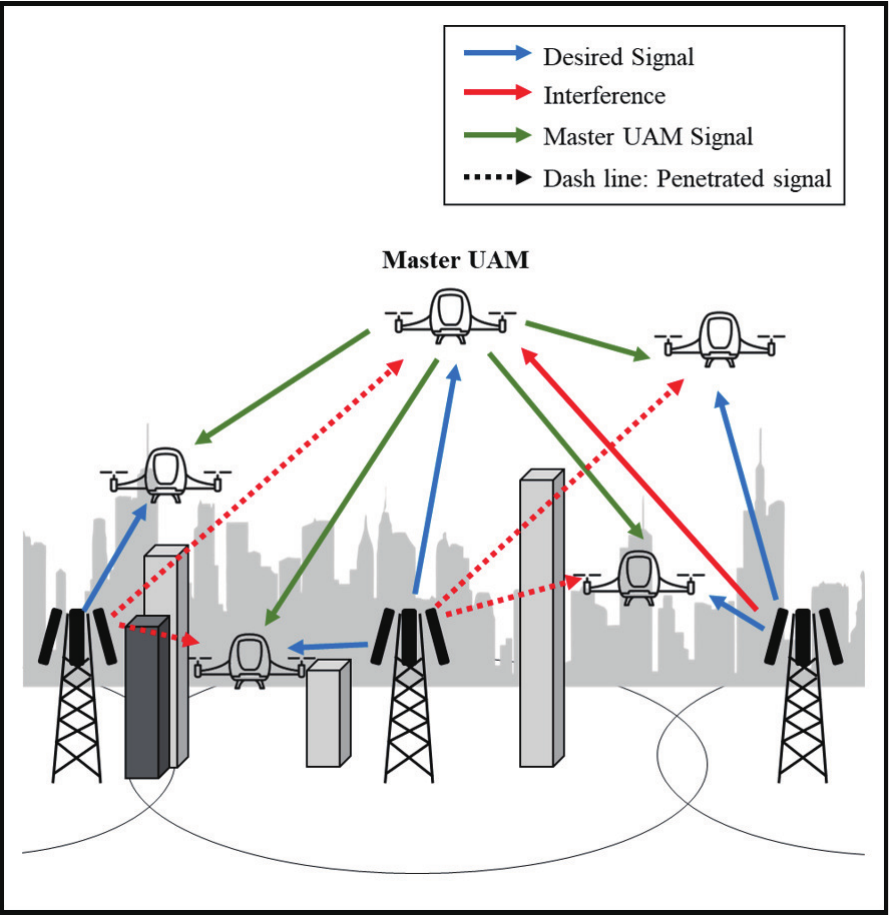}
    \caption{The concept assistive transmission of the master UAM.}
    \label{Fig4_11}
\end{figure}

\clearpage

    \item \textbf{Master UAM Selection}
    
    \,\,\,\,Denote that ${\Phi _{U_F}}$ and ${\Phi _{U_S}}$ are the set of failure and success UAMs to transmit the target data rate $\tau_c$, respectively. We define $D_j$ as the deficiency of data rate required for $j_{th}$ UAM as $D_j = \tau_j - \tau_c.$
    Positive deficiency means that the target data rate is satisfied and there is a spare amount of resources, and a negative value means that the target rate is not satisfied.
    
    \,\,\,\,Our objective is to select a UAM as a master UAM of which deficiency is larger than the sum of deficiency in ${\Phi _{U_F}}$. The detailed algorithm is as follows:

    \begin{algorithm}
   \caption{Master UAM Selection Algorithm} 
   \begin{algorithmic}[1]
   \State {\bf{Input:}} UAM Rate $\tau_j$ and target rate $\tau_c$
   \State {\bf{Output:}} $m$, the index of master UAM \\
    \,\,\,\,\,\,\,\,\,\,\,\,\,\,\,\,\,\,\,\,\,\,
    ${\Phi _{U_F}}$, the set of failure UAMs failure UAMs not to success $\tau_c$
   \State {\bf{Initialize:}} $D_j$, the deficiency of data rate required for $j_{th}$ UAM \\
    \,\,\,\,\,\,\,\,\,\,\,\,\,\,\,\,\,\,\,\,\,\,\,\,\,\,
    $D_{max} \leftarrow 0$, the maximum value of $D_j$ \\
    \,\,\,\,\,\,\,\,\,\,\,\,\,\,\,\,\,\,\,\,\,\,\,\,\,\,
   ${\Phi _{U_S}}$, the set of success UAMs to transmit $\tau_c$ 
   \ForAll {$j$ $\in$ ${\Phi _U}$}
      \State $D_j = \tau_j - \tau_c$
   \If {$D_j < 0$ }
   \State Include $j$ into ${\Phi _{U_F}}$
   \Else 
   \State Include $j$ into ${\Phi _{U_S}}$
   \EndIf
   \EndFor
   \ForAll{$j$ $\in$ ${\Phi _{U_S}}$}
   \If {$D_j > \sum\limits_{p \in {\Phi _{{U_F}}}}^{} { - {D_p}} $ \& $D_j>D_{max}$ }
   \State $D_{max} = D_j$
   \State $m = j$
   \EndIf
   \EndFor
   \end{algorithmic} 
\end{algorithm}

\clearpage

\begin{figure}[h!]
    \centering
    \includegraphics[width=0.9\columnwidth]{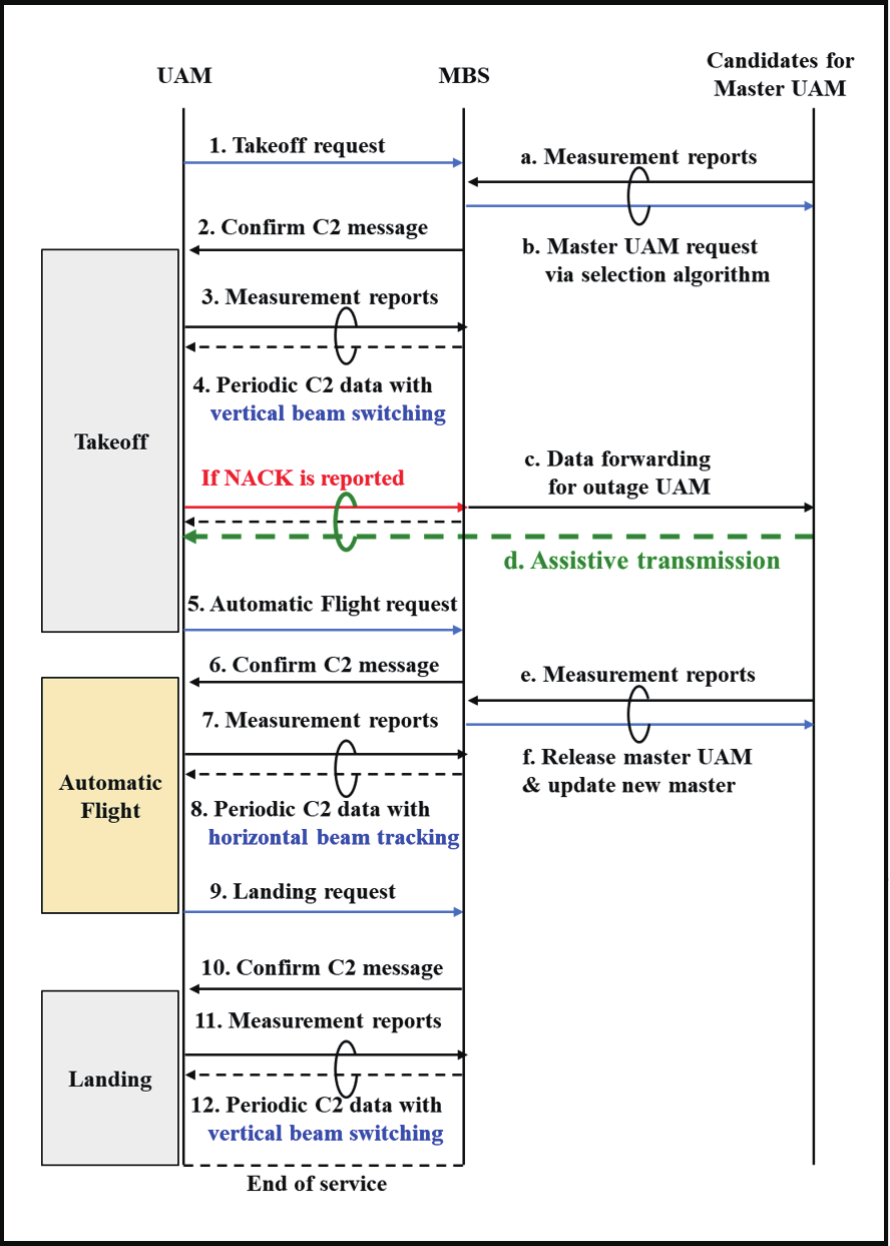}
    \caption{The example of the signal flow for the proposed UAM networks.}
    \label{Fig4_12}
\end{figure}

\clearpage

\item \textbf{The Final Result of UAM Networks with Master UAM Transmissions}

\,\,\,\,Figure \ref{Fig4_13} shows the reliability performance of each UAM with assistive transmissoins. We assume that all UAMs have a single Omni-directional antenna with transmit power $23$ $dBm$. The dedicated bandwidth for broadcasting of the master UAM is set to 5 MHz. We prove that the mater UAM can operate a stable UAN network by additionally sending a C2 message for outage UAM placed in the shadow area in cooperation with MBS. Not only do most UAMs satisfy the reliability requirement compared to Figure \ref{Fig4_9} (b), but the overall average reliability performance of the system also shows more than 99.9 \%.

\begin{figure}[h!]
    \centering
    \includegraphics[width=0.8\columnwidth]{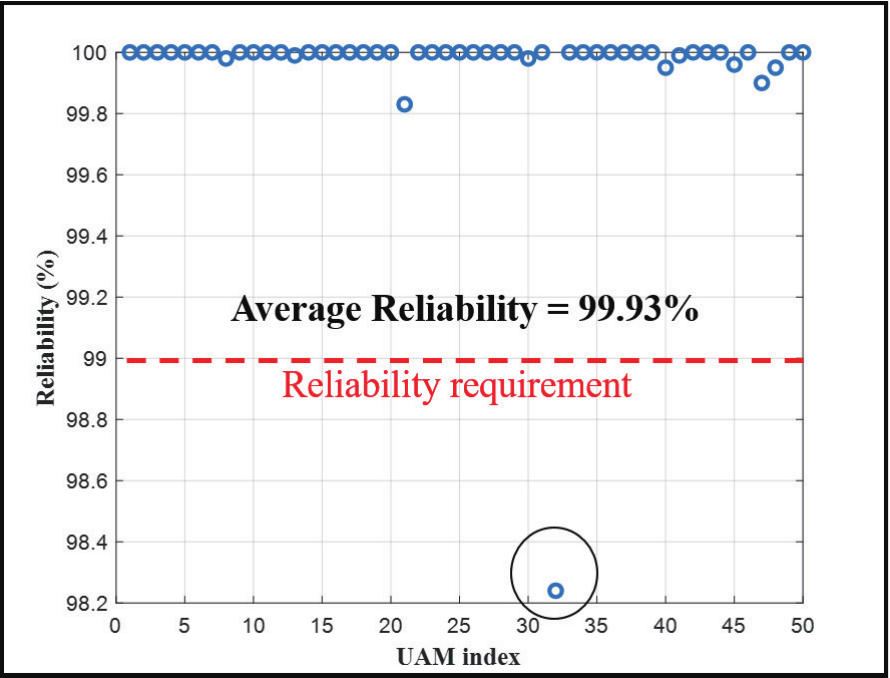}
    \caption{The final result of the UAMs' reliability performance.}
    \label{Fig4_13}
\end{figure}

\end{itemize}

\chapter{Conclusion}\label{Chap_5}
\hspace{-5mm}\rule{135mm}{1pt}

{5.1 Dissertation Summary 

5.2 Further Research Issues\\}
\rule{135mm}{1pt}
\clearpage

\section{Dissertation Summary}

We investigated the performance of UAM networks and proposed a design method to achieve 99.9 \% reliability. First, we analyzed the blockage effect of the urban buildings scenario through stochastic geometry model. The influence of the shape, height, and density of the building on the communication system was investigated, and through this, an appropriate MBS configuration was discussed.

Using the previously obtained blockage probability model, the penetration losses of the desired signal and the interference signal were analyzed. It was verified that the proper density and height of the building blocks interfering signals and plays a positive role in enhancing the quality of the received signal. We also derived the coverage probability performance reflecting the penetration loss by the building. From the experimental observation, the performance of each UAM height was demonstrated, and the characteristics of each performance area were examined.

Through the performance analysis, we identified cell edge area at high altitudes, and tried to overcome the area by proposing three methods. First, we adjusted the optimal tilting antenna angle targeting to the cell edge UAMs. Second, through the multi-layered narrow beam as well as the frequency reuse pattern, the power of the desired signal was increased and the amount of interference was reduced. Lastly, by defining UAM with margin of data transmission as master, it supported outage UAMs, and finally, 99.9 \% reliability were achieved. In conclusion, differentiating the conventional terrestrial wireless communications, we proposed a specialized architecture for ultra-reliable UAM networks that can achieve the target reliability defined in the 3GPP.

\section{Further Research Issues}
For further researches, some interesting issues are summarized as follows.

\begin{itemize}

\item {\textbf{How to handle control message within handover periods}}
    
    \,\,\,\,It is necessary to analyze the effect of handover procedure time on UAM communication performance. Usually, the time required for the handover procedure is about 50ms. In the case of receiving an urgent C2 message during the handover process, it may be difficult to satisfy the required reliability. Since UAM's route information can be known in advance, we can consider sequential handover based on the information.

\item {\textbf{How to reduce the delay time in master UAM transmissions}}

    \,\,\,\,In our study, we did not consider the delay occurring in the process of selecting master UAM and its assistive transmissions. However, because of the fast-moving UAM, a small delay can cause the information used to select the master UAM to be outdated, which can degrade the transmission efficiency. In this sense, a study on the master UAM selection technique considering mobility is necessary.

\item {\textbf{How to manage the interference among multiple master UAMs}}

    \,\,\,\,In the 3D urban simulator, it was assumed that there is only one mater UAM in the observation region. However, when viewed over a large area, multiple master UAMs may exist, and interference control among these UAMs is required.

\end{itemize}


\begin{thebibliography}{10}
\providecommand{\url}[1]{#1}
\csname url@samestyle\endcsname
\providecommand{\newblock}{\relax}
\providecommand{\bibinfo}[2]{#2}
\providecommand{\BIBentrySTDinterwordspacing}{\spaceskip=0pt\relax}
\providecommand{\BIBentryALTinterwordstretchfactor}{4}
\providecommand{\BIBentryALTinterwordspacing}{\spaceskip=\fontdimen2\font plus
\BIBentryALTinterwordstretchfactor\fontdimen3\font minus \fontdimen4\font\relax}
\providecommand{\BIBforeignlanguage}[2]{{%
\expandafter\ifx\csname l@#1\endcsname\relax
\typeout{** WARNING: IEEEtran.bst: No hyphenation pattern has been}%
\typeout{** loaded for the language `#1'. Using the pattern for}%
\typeout{** the default language instead.}%
\else
\language=\csname l@#1\endcsname
\fi
#2}}
\providecommand{\BIBdecl}{\relax}
\BIBdecl

\bibitem{0}
\textrm{NASA Urban Air Mobility}, ``\textrm{URBAN AIR MOBILITY (UAM) MARKET STUDY (URL: https://www.nasa.gov/sites/default/files/atoms/files/uam-market-study-executive-summary-v2.pdf)},'' Nov. 2018.

\bibitem{1}
\textrm{3GPP TS 22.125. v17.2.0}, ``\textrm{3rd Generation Partnership Project; Technical Specification Group Services and System Aspects; Unmanned Aerial System (UAS) support in 3GPP; Stage 1 (Release 17)},'' Sept. 2020.

\bibitem{3}
T.~{Izydorczyk}, G.~{Berardinelli}, P.~{Mogensen}, M.~M. {Ginard}, J.~{Wigard}, and I.~Z. {Kovács}, ``Achieving high uav uplink throughput by using beamforming on board,'' \emph{IEEE Access}, vol.~8, pp. 82\,528--82\,538, 2020.

\bibitem{4}
W.~{Mei}, Q.~{Wu}, and R.~{Zhang}, ``Cellular-connected uav: Uplink association, power control and interference coordination,'' \emph{IEEE Transactions on Wireless Communications}, vol.~18, no.~11, pp. 5380--5393, 2019.

\bibitem{5}
W.~{Mei} and R.~{Zhang}, ``Uplink cooperative noma for cellular-connected uav,'' \emph{IEEE Journal of Selected Topics in Signal Processing}, vol.~13, no.~3, pp. 644--656, 2019.

\bibitem{6}
N.~{Senadhira}, S.~{Durrani}, X.~{Zhou}, N.~{Yang}, and M.~{Ding}, ``Uplink noma for cellular-connected uav: Impact of uav trajectories and altitude,'' \emph{IEEE Transactions on Communications}, vol.~68, no.~8, pp. 5242--5258, 2020.

\bibitem{7}
J.~{Lu}, Y.~{Wang}, T.~{Liu}, Z.~{Zhuang}, X.~{Zhou}, F.~{Shu}, and Z.~{Han}, ``Uav-enabled uplink non-orthogonal multiple access system: Joint deployment and power control,'' \emph{IEEE Transactions on Vehicular Technology}, vol.~69, no.~9, pp. 10\,090--10\,102, 2020.

\bibitem{8}
R.~{Duan}, J.~{Wang}, C.~{Jiang}, H.~{Yao}, Y.~{Ren}, and Y.~{Qian}, ``Resource allocation for multi-uav aided iot noma uplink transmission systems,'' \emph{IEEE Internet of Things Journal}, vol.~6, no.~4, pp. 7025--7037, 2019.

\bibitem{9}
J.~{Seo}, S.~{Pack}, and H.~{Jin}, ``Uplink noma random access for uav-assisted communications,'' \emph{IEEE Transactions on Vehicular Technology}, vol.~68, no.~8, pp. 8289--8293, 2019.

\bibitem{10}
S.~{Zhang}, H.~{Zhang}, B.~{Di}, and L.~{Song}, ``Cellular uav-to-x communications: Design and optimization for multi-uav networks,'' \emph{IEEE Transactions on Wireless Communications}, vol.~18, no.~2, pp. 1346--1359, 2019.

\bibitem{11}
W.~{Mei} and R.~{Zhang}, ``Cooperative downlink interference transmission and cancellation for cellular-connected uav: A divide-and-conquer approach,'' \emph{IEEE Transactions on Communications}, vol.~68, no.~2, pp. 1297--1311, 2020.

\bibitem{12}
L.~{Liu}, S.~{Zhang}, and R.~{Zhang}, ``Multi-beam uav communication in cellular uplink: Cooperative interference cancellation and sum-rate maximization,'' \emph{IEEE Transactions on Wireless Communications}, vol.~18, no.~10, pp. 4679--4691, 2019.

\bibitem{13}
Z.~{Li}, Y.~{Wang}, M.~{Liu}, R.~{Sun}, Y.~{Chen}, J.~{Yuan}, and J.~{Li}, ``Energy efficient resource allocation for uav-assisted space-air-ground internet of remote things networks,'' \emph{IEEE Access}, vol.~7, pp. 145\,348--145\,362, 2019.

\bibitem{14}
C.~{Joo} and J.~{Choi}, ``Low-delay broadband satellite communications with high-altitude unmanned aerial vehicles,'' \emph{Journal of Communications and Networks}, vol.~20, no.~1, pp. 102--108, 2018.

\bibitem{15}
D.~{Yang}, Q.~{Wu}, Y.~{Zeng}, and R.~{Zhang}, ``Energy tradeoff in ground-to-uav communication via trajectory design,'' \emph{IEEE Transactions on Vehicular Technology}, vol.~67, no.~7, pp. 6721--6726, 2018.

\bibitem{16}
M.~{Hua}, L.~{Yang}, Q.~{Wu}, and A.~L. {Swindlehurst}, ``3d uav trajectory and communication design for simultaneous uplink and downlink transmission,'' \emph{IEEE Transactions on Communications}, vol.~68, no.~9, pp. 5908--5923, 2020.

\bibitem{17}
S.~{Zhang}, Y.~{Zeng}, and R.~{Zhang}, ``Cellular-enabled uav communication: A connectivity-constrained trajectory optimization perspective,'' \emph{IEEE Transactions on Communications}, vol.~67, no.~3, pp. 2580--2604, 2019.

\bibitem{18}
F.~{Jiang} and A.~L. {Swindlehurst}, ``Optimization of uav heading for the ground-to-air uplink,'' \emph{IEEE Journal on Selected Areas in Communications}, vol.~30, no.~5, pp. 993--1005, 2012.

\bibitem{19}
M.~A. {Ali} and A.~{Jamalipour}, ``Uav placement and power allocation in uplink and downlink operations of cellular network,'' \emph{IEEE Transactions on Communications}, vol.~68, no.~7, pp. 4383--4393, 2020.

\bibitem{20}
A.~{Ranjha} and G.~{Kaddoum}, ``Quasi-optimization of uplink power for enabling green urllc in mobile uav-assisted iot networks: A perturbation-based approach,'' \emph{IEEE Internet of Things Journal}, vol.~8, no.~3, pp. 1674--1686, 2021.

\bibitem{21}
M.~{Mozaffari}, A.~{Taleb Zadeh Kasgari}, W.~{Saad}, M.~{Bennis}, and M.~{Debbah}, ``Beyond 5g with uavs: Foundations of a 3d wireless cellular network,'' \emph{IEEE Transactions on Wireless Communications}, vol.~18, no.~1, pp. 357--372, 2019.

\bibitem{22}
L.~{Bai}, R.~{Han}, J.~{Liu}, Q.~{Yu}, J.~{Choi}, and W.~{Zhang}, ``Air-to-ground wireless links for high-speed uavs,'' \emph{IEEE Journal on Selected Areas in Communications}, vol.~38, no.~12, pp. 2918--2930, 2020.

\bibitem{24}
W.~{Tang}, H.~{Zhang}, and Y.~{He}, ``Tractable modelling and performance analysis of uav networks with 3d blockage effects,'' \emph{IEEE Wireless Communications Letters}, vol.~9, no.~12, pp. 2064--2067, 2020.

\bibitem{25}
T.~{Bai}, R.~{Vaze}, and R.~W. {Heath}, ``Analysis of blockage effects on urban cellular networks,'' \emph{IEEE Transactions on Wireless Communications}, vol.~13, no.~9, pp. 5070--5083, 2014.

\bibitem{26}
J.~G. {Andrews}, F.~{Baccelli}, and R.~K. {Ganti}, ``A tractable approach to coverage and rate in cellular networks,'' \emph{IEEE Transactions on Communications}, vol.~59, no.~11, pp. 3122--3134, 2011.

\bibitem{27}
\textrm{3GPP TR 36.873. v12.7.0}, ``\textrm{3rd Generation Partnership Project; Technical Specification Group Radio Access Network; Study on 3D channel model for LTE (Release 12)},'' Dec. 2017.

\bibitem{28}
S.~{Sun}, T.~S. {Rappaport}, S.~{Rangan}, T.~A. {Thomas}, A.~{Ghosh}, I.~Z. {Kovacs}, I.~{Rodriguez}, O.~{Koymen}, A.~{Partyka}, and J.~{Jarvelainen}, ``Propagation path loss models for 5g urban micro- and macro-cellular scenarios,'' in \emph{2016 IEEE 83rd Vehicular Technology Conference (VTC Spring)}, 2016, pp. 1--6.

\end{thebibliography}
\end{document}